\documentclass[12pt,preprint]{aastex}

\def\logz{\lbrack\hbox{Fe/H}\rbrack}

\shorttitle{Sextans A Cepheids}
\shortauthors{Dolphin et al.}

\begin{document}

\title{Deep HST Imaging of Sextans A. II. Cepheids and Distance
\linespread{1.0}\footnote{Based on observations with the NASA/ESA \textit{Hubble Space Telescope}, obtained at the Space Telescope Science Institute, which is operated by the Association of Universities for Research in Astronomy, Inc., under NASA contract NAS 5-26555. These observations are associated with proposal ID 7496.}}

\author{Andrew E. Dolphin, A. Saha}
\affil{Kitt Peak National Observatory, National Optical Astronomy Observatories, P.O. Box 26372, Tucson, AZ 85726}
\email{dolphin@noao.edu, saha@noao.edu}

\author{Evan D. Skillman, R.C. Dohm-Palmer}
\affil{Astronomy Department, University of Minnesota, Minneapolis, MN 55455}
\email{skillman@astro.umn.edu, rdpalmer@astro.umn.edu}

\author{Eline Tolstoy, A.A. Cole}
\affil{Kapteyn Institute, University of Groningen, P.O. Box 800, 9700AV Groningen, the Netherlands}
\email{etolstoy@astro.rug.nl, cole@astro.rug.nl}

\author{J.S. Gallagher, J.G. Hoessel}
\affil{University of Wisconsin, Dept. of Astronomy, 475 N. Charter St, Madison, WI 53706}
\email{jsg@astro.wisc.edu, hoessel@astro.wisc.edu}

\and

\author{Mario Mateo}
\affil{University of Michigan, Dept. of Astronomy, 821 Dennison Building, Ann Arbor, MI 48109-1090}
\email{mateo@astro.lsa.umich.edu}

\begin{abstract}
We have identified 82 short-period variable stars in Sextans A from deep WFPC2 observations.  All of the periodic variables appear to be short-period Cepheids, with periods as small as 0.8 days for fundamental-mode Cepheids and 0.5 days for first-overtone Cepheids.  These objects have been used, along with measurements of the RGB tip and red clump, to measure a true distance modulus to Sextans A of $\mu_0 = 25.61 \pm 0.07$, corresponding to a distance of $d = 1.32 \pm 0.04$ Mpc.  Comparing distances calculated by these techniques, we find that short-period Cepheids ($P < 2$ days) are accurate distance indicators for objects at or below the metallicity of the SMC.  As these objects are quite numerous in low-metallicity star-forming galaxies, they have the potential for providing extremely precise distances throughout the Local Group.  We have also compared the relative distances produced by other distance indicators.  We conclude that calibrations of RR Lyraes, the RGB tip, and the red clump are self-consistent, but that there appears to be a small dependence of long-period Cepheid distances on metallicity.  Finally, we present relative distances of Sextans A, Leo A, IC 1613, and the Magellanic Clouds.
\end{abstract}


\section{Introduction}

Sextans A (DDO 75) is a dwarf irregular galaxy, which lies in a small group of galaxies including NGC 3109, Sextans B, and the Antlia dwarf at a distance of approximately $1.4$ Mpc.  The group appears to not be bound to the Local Group \citep{van99}, making it the nearest group of galaxies to us.

Cepheids were discovered in the galaxy by \citet{san82}, who calculated a distance modulus of $\mu_0 = 25.6 \pm 0.2$.  Later work \citep{san85,wal87} improved the photometry and adjusted the Cepheid magnitudes by over half a magnitude fainter, a change that was largely counteracted by revisions to the period-luminosity relation.  \citet{pio94} reobserved the five Sandage \& Carlson Cepheids and added observations of five new Cepheids in Sextans A.  They derived a distance modulus of $\mu_0 = 25.71 \pm 0.20$ based on a distance modulus of 18.5 for the LMC.  \citet{sak96} added new, single epoch photometry of the six Cepheids with periods between 10 and 25 days and calculated a distance modulus of $\mu_0 = 25.85 \pm 0.15$ as well as an RGB tip distance modulus of $\mu_0 = 25.74 \pm 0.13$.

The distance to Sextans A is important for three reasons.  First, as noted above, the galaxy is part of the Antlia-Sextans group, which may or may not be bound to the Local Group.  Improved measurements of the distances of the four members should help to clarify this issue.  Second, accurate distances to all of the nearest galaxies are important for an accurate characterization of the local mass distribution and the degree to which galaxy distributions follow mass distributions (e.g., Peebles et al. 2001, Ekholm et al. 2001, and references therein).  Perhaps of greatest interest is that Sextans A is one of the lowest metallicity objects in which classical Cepheids have been discovered (with HII region abundances of 12 + log[O/H] $\approx$ 7.6 Skillman, Kennicutt, \& Hodge 1989; Pilyugin 2001).  Thus, observationally studies of the variable stars in Sextans A can shed light on the question of the metallicity dependence of the Cepheid distance scale (e.g., Kennicutt et al.\ 1998).

HST observations permit a more accurate measurement of the distance of Sextans A.  Not only does the increased resolution permit more accurate photometry of the brighter distance indicators such as the RGB tip, but the deeper photometry allows the red clump position to be measured for the first time.  Additionally, our study of Leo A \citep{dol02} indicates that we should find large numbers of short-period Cepheids in low-metallicity star-forming galaxies (such as Sextans A); accurate photometry of these objects is difficult from the ground but lies easily within the capabilities of HST and can potentially provide a new distance measurement.

This paper is the second in our study of Sextans A based on deep WFPC2 images.  In Paper I \citep{doh02}, we used the blue helium-burning (BHeB) stars to examine its recent star formation history.  In this paper, we use the 3-day observation baseline to examine the variable star population.  Section 2 describes our photometry and identification of variables.  In Section 3, we examine the short-period Cepheid population and determine the distance to Sextans A.  Section 4 is a comparison of distance moduli to Sextans A and other dwarf galaxies, as determined using a variety of standard candles.

\section{Data and Reduction}

\subsection{Observations}

The observations were described in detail in Paper I.  Summarizing briefly, the data consist of 24 pairs of 1200 second images.  Sixteen of the pairs were taken through the F814W filter, eight through F555W.  Data were taken with four dithering positions, with one quarter of the observations (four pair of F814W images and two pair of F555W images) made at each.  A pair of F656N images was also taken, but was not used for our variable star work and thus will not be referenced again in this paper.  The total baseline of observations was slightly over 3 days.

\subsection{Photometry}

The data were obtained from the STScI archive using on the fly calibration, and thus were pipeline-calibrated using the best available calibration images at the time of retrieval.  Reduction and photometry was made with the HSTphot package \citep{dol00a}.  For this variable star study, each pair of images was cosmic ray cleaned and combined, to give effective 2400 second images.  Additionally, a deep F555W frame at the first pointing was created to provide a reference image for alignment.  The mosaic of this image is shown in Figure \ref{figImage}.
\placefigure{figImage}

Photometry was obtained on the 24 images simultaneously, producing photometry for all epochs for a common list of stars.  Calibration to the WFPC2 flight system was made using HSTphot-produced CTE corrections and calibrations \citep{dol00b}.  Figure \ref{figCMD} shows the color-magnitude diagram (CMD) for all well-photometered stars (those with $\chi \le 2.0$, $|$sharpness$| \le 0.3$, total signal-to-noise of at least 5, and observed in both filters).
\placefigure{figCMD}

\subsection{Variable Star Identification}

Variable star candidates were identified first using a series of automated cuts, similar to those used in our study of IC 1613 \citep{dol01a}.  The star had to meet each of the following five criteria, testing the photometry, variability, and periodicity.

First, the object had to be a well-photometered star.  While it is possible for a variable star to be part of a blend, our PSF-fitting photometry attempts to fit the profile to that of a single star and thus is unreliable for photometering blends.

Our next cuts were intended to eliminate non-variable or weakly-variable stars.  The simplest requirement was that the rms magnitude scatter,
\begin{equation}
\sigma_{mag} = \sqrt{ \frac{1}{N_{F555W}+N_{F814W}} ( \sum_{i=1}^{N_{F555W}} (F555W_i - \overline{F555W})^2 + \sum_{i=1}^{N_{F814W}} (F814W_i - \overline{F814W})^2 ) }
\end{equation}
had to be 0.14 magnitudes or greater.  Next, the overall reduced $\chi^2$,
\begin{equation}
\chi^2 = \frac{1}{N_{F555W}+N_{F814W}} ( \sum_{i=1}^{N_{F555W}} \frac{(F555W_i - \overline{F555W})^2}{\sigma_i^2} + \sum_{i=1}^{N_{F814W}} \frac{(F814W_i - \overline{F814W})^2}{\sigma_i^2} )
\end{equation}
had to be 3.0 or greater.  These two cuts eliminate stars that are weakly variable and those for which the photometry signal-to-noise is insufficient to discern variability.  We also clipped 1/3 of the extreme points and required that the recalculated reduced $\chi^2$ exceed 0.5; this was intended to eliminate non-variable stars with a few bad points.

The final test was a modified Lafler-Kinman algorithm \citep{laf65}, which tests for periodicity.  This was implemented by computing $\Theta$ for periods between 0.1 and 4.0 days.  The $\Theta$ parameter is calculated by determining the light curve for a trial period and using the equation
\begin{equation}
\Theta = \frac {\sum_{i=1}^{N}(m_i - m_{i+1})^2}{\sum_{i=1}^{N}(m_i - \overline{m})^2},
\end{equation}
where $N$ is the number of exposures for a given filter, $m_i$ is the magnitude at phase $i$, and $\overline{m}$ is the mean magnitude.  If the trial period is the correct period, each magnitude $m_i$ will be close to the adjacent magnitude $m_{i+1}$, giving a value of $\Theta$ that decreases as $1/N^2$.  If the trial period is incorrect, there will be less correlation and consequently larger values of $\Theta$.  Because we had data in two filters, we combined the $\Theta$ values with:
\begin{equation}
\Theta = \frac{4}{ ( 1/\sqrt{\Theta_{F555W}} + 1/\sqrt{\Theta_{F814W}} )^2}.
\end{equation}
For a star to pass the periodicity test, its minimum value of $\Theta$ had to be 0.85 or less.

Our photometry identified 39947 total objects, of which 33681 were well-photometered stars and 110 passed our automatic selection process.  Nine of these stars appeared to not be clean detections upon manual examination of the images and were removed from the list.  Each of the remaining stars was examined manually and subjectively graded based on its quality.  A quality value of 4 means that the variable had clean light curves in both filters, the F555W and F814W light curves were in phase, and the period had no ambiguity.  Lower quality values were assigned to variables that failed to meet one or more of these criteria.  Only those variables with quality of 3 or 4 were kept; 82 passed this test.

To compute mean magnitudes, we computed a phase-weighted average using
\begin{equation}
\langle m \rangle = -2.5 \log \sum_{i=1}^{N} \frac{\phi_{i+1} - \phi_{i-1}}{2} 10^{-0.4 m_i},
\end{equation}
where $\phi_i$ is the phase and $m_i$ is the magnitude at each point along the light curve.  These values, still in WFPC2 F555W and F814W magnitudes, were then transformed to standard $V$ and $I$ by application of color terms.  Complete photometry for all variable stars is shown in Table \ref{tab_phot} and summarized in Table \ref{tabvars}; light curves are shown in Figure \ref{figLCs}.
\placetable{tab_phot}
\placetable{tabvars}
\placefigure{figLCs}

Although location on the CMD was not a criterion, all of our variable stars were found near the expected location of short-period Cepheids -- brighter and bluer than the red clump.  Figure \ref{figVCMD} shows the Sextans A CMD, with the variable stars highlighted.  The apparent horizontal extensions of the Cepheid population at $I \approx 23.9$ are artifacts of poor phase coverage in F555W, explained in the next section.

It should be emphasized that our detection efficiency was not 100\%.  Thus we cannot rule out the existence of Population II Cepheids, nor are we confident that the ``non-variable'' stars falling within the instability strip are not, in fact, variables that were not detected.  Additionally, while the CMD from combined photometry is 50\% complete to $I = 27$, the single epoch signal-to-noise of stars below $I=25$ is such that detection of fainter periodic variables, such as RR Lyraes, would have been extremely difficult.  For example, the signal-to-noise of our individual epoch photometry at the expected location of RR Lyraes is $\sim 5$ in $I$; thus a variable at that magnitude would have to have an rms variability of $\sim 0.5$ magnitudes in $I$ to pass all of our variability tests.
\placefigure{figVCMD}

\subsection{Phase Coverage}

\begin{figure}
\plotone{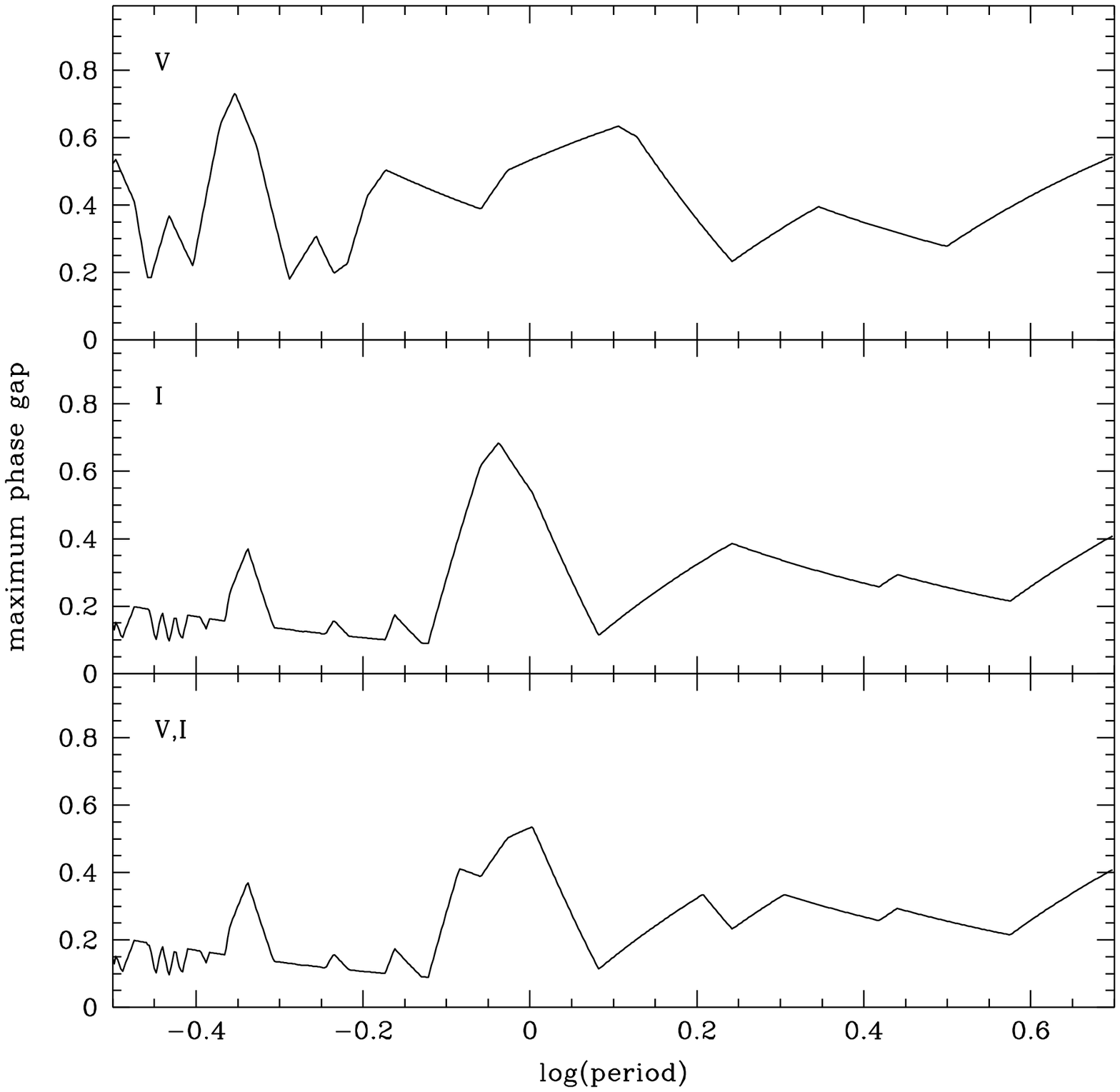}
\caption{Maximum gap in phase coverage as a function of period.  The top panel shows the gap in F555W, the middle panel in F814W, and the bottom panel the lesser of the two. \label{figdphase}}
\end{figure}

Since the primary goal of these observations was the obtaining of deep photometry, the program was not carried out in a way that would maximize the phase coverage.  Thus, a variable star with one period may have excellent coverage of its light curve while one with another period may have poor coverage.  Figure \ref{figdphase} shows the maximum gap in our phase coverage as a function of period, assuming that the star was measured at every epoch.  While the F814W observations consist of 16 epochs and thus give the better coverage, we will have poor phase coverage of any variable whose period is between 0.8 and 1.1 days.  For such a star, all 16 F814W observations would cover only about 40\% of the light curve, while the 8 F555W observations cover about 50\%.  Thus our detection efficiency will suffer, since such a star may not show sufficient variability over the phase range included in our observations to trigger our $\chi^2$ flag.  Additionally, if such a variable is recovered, the absence of half the light curve prevents an accurate mean magnitude determination and creates ambiguity between periods of a day and half a day.  Fortunately, the color of such a star should be measured acceptably since both light curves would have their gaps at close to the same phases.
\placefigure{figdphase}

In addition to detection efficiency, poor phase coverage affects our period and mean magnitude determinations.  Eight stars could have been fit with either one-day or half-day light curves; fortunately all such stars could be identified using their locations in the period-luminosity diagram.  The mean magnitude errors are more problematic, since the lack of half the light curve can bias the calculation in either direction if the star is observed only near maximum or minimum brightness.  For stars with periods of about one day, this affected the mean magnitudes in both filters similarly, since the same phases were observed in both filters.  However, for stars with periods of about 1.2 days, the phase coverage in F814W was excellent (maximum phase gap of 0.1) while that in F555W was poor (maximum phase gap of 0.6).  Five such stars were observed only near minimum in F555W; these are the five that appear to be on top of the red clump in Figure \ref{figVCMD} ($I \simeq 23.9$ and $V-I \simeq 0.8$).  Correcting for this measurement error, we find that the five are, on average, 0.4 magnitudes brighter in $V$ (and thus bluer) than our measurement.  Similarly, four 1.2-day variables were observed exclusively near maximum in F555W, creating the apparent blue extension at the same $I$ magnitude.

\section{Analysis}

\subsection{Cepheids}

\begin{figure}
\plotone{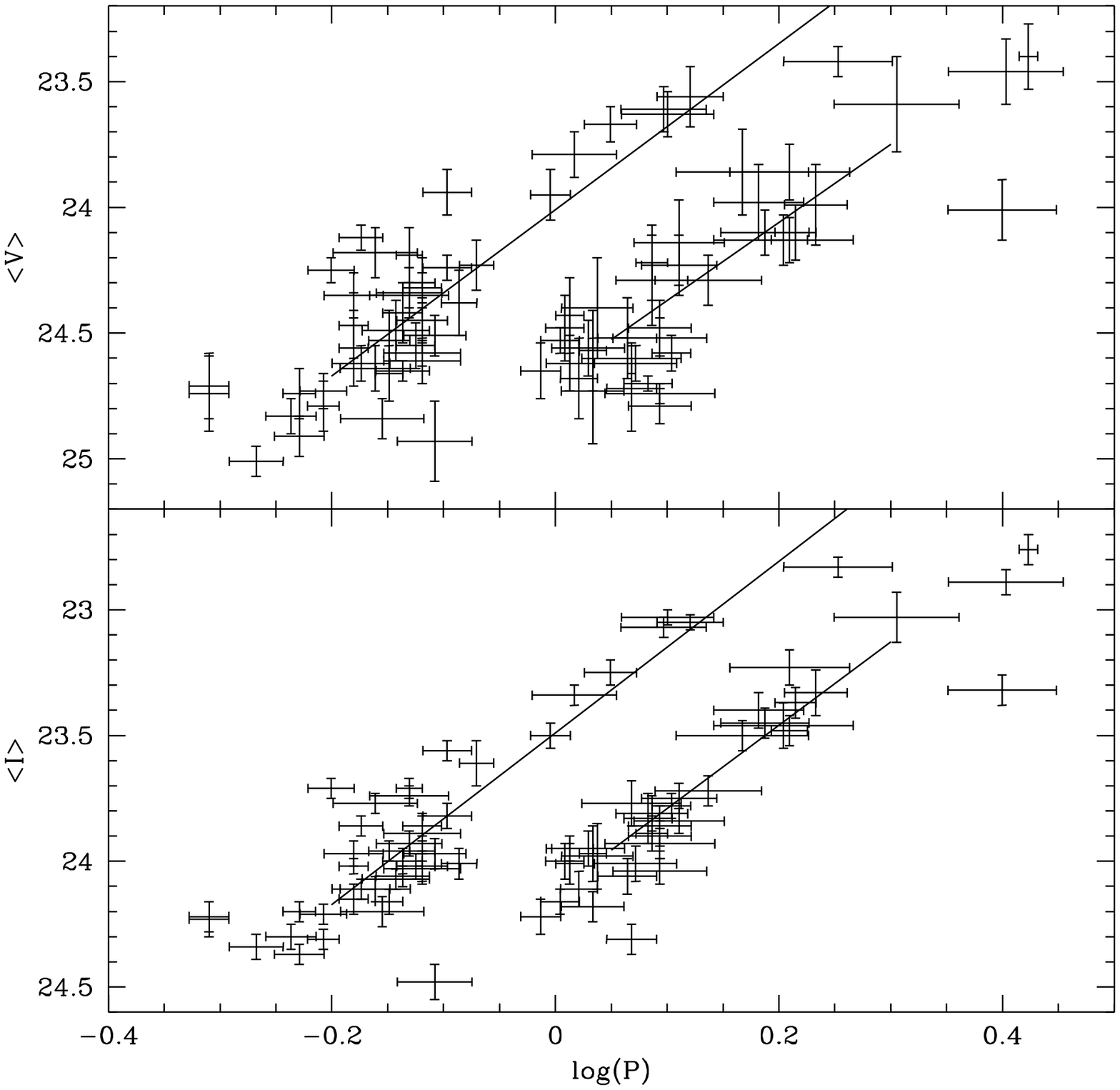}
\caption{Period-luminosity relations in $V$ and $I$ for short-period Cepheids in Sextans A.  Slopes are calculated from the OGLE SMC Cepheid database \citep{uda99a}.  In both panels, the first overtone P-L relation is the upper line and the fundamental mode relation is the lower line. \label{figPL}}
\end{figure}

Figure \ref{figPL} shows our period-luminosity (P-L) relations in $V$ and $I$.  The $I$ relation is significantly sharper for two reasons -- better phase coverage in F814W allows for a better measurement of the mean magnitude, and the P-L relation has less scatter in redder bands.  We note that one of the fundamental-mode pulsators (C4-V24) has a period of only $0.78 \pm 0.06$ days; such objects were also seen in our variable star study of Leo A \citet{dol02}.  We note that the gap between C4-V24 and the one with the next-shortest period (C1-V07; $P = 0.97 \pm 0.04$ days) is the result of our poor phase coverage in that period range as described in the previous section.  As with the three $\sim 0.8$ day Cepheids in Leo A, we believe C4-V24 to be a \textit{bona fide} classical Cepheid rather than an RR Lyrae or anomalous Cepheid. 
\placefigure{figPL}

Because the P-L relations have have steeper slopes at short periods than at long periods \citep{bau99}, we calculated relations using short-period ($< 2$ day) Cepheids (638 fundamental-mode and 611 first-overtone pulsators) from the OGLE SMC Cepheid database \citep{uda99a}:
\begin{equation}
M_V (\hbox{FM}) = -3.10 \log(P) - 1.04 \pm 0.01
\end{equation}
\begin{equation}
M_I (\hbox{FM}) = -3.31 \log(P) - 1.56 \pm 0.01
\end{equation}
\begin{equation}
M_V (\hbox{FO}) = -3.30 \log(P) - 1.70 \pm 0.01
\end{equation}
\begin{equation}
M_I (\hbox{FO}) = -3.41 \log(P) - 2.17 \pm 0.01
\end{equation}
assuming an SMC distance modulus of $\mu_0 = 18.88$ \citet{dol01b}.

Converting to the reddening-free parameter $W \equiv 2.45 I - 1.45 V$, the relations become
\begin{equation}
M_W (\hbox{FM}) = -3.61 \log(P) - 2.30
\end{equation}
\begin{equation}
M_W (\hbox{FO}) = -3.57 \log(P) - 2.86
\end{equation}
The true distance modulus is thus given by $W - M_W$, allowing us to measure a distance for each Sextans A Cepheid.  Averaging the individual distances of our Cepheids, we measure true distance moduli of $\mu_0 = \mu_0(SMC) + 6.76 \pm 0.05$ from 39 fundamental-mode Cepheids and $\mu_0 = \mu_0(SMC) + 6.70 \pm 0.03$ from 43 first-overtone Cepheids.  For the complete sample of 82 Cepheids, we measure a true distance modulus of $\mu_0 = \mu_0(SMC) + 6.72 \pm 0.03 = 25.60 \pm 0.09$.

Additionally, we measure a mean extinction of $A_V = mu_V - mu_0 = 0.12 \pm 0.04$ ($A_I = 0.07 \pm 0.03$).  The agreement with the value of $A_V = 0.14$ measured from the maps of \citet{sch98} supports our estimate of the extinction necessary for the distance measurement.  Fitting P-L relations (slopes and zero-points) to the Sextans A data, we find that the slopes are consistent with those of the SMC data; this is also provides confirmation of the reliability of our application of the SMC relation to Sextans A.  We will discuss this more in section 4.

The lines in Figure \ref{figPL} are the above relations, shifted to our measured apparent distance moduli of $\mu_V = 25.72$ and $\mu_I = 25.67$.

\subsection{The Distance to Sextans A}

\begin{figure}
\plotone{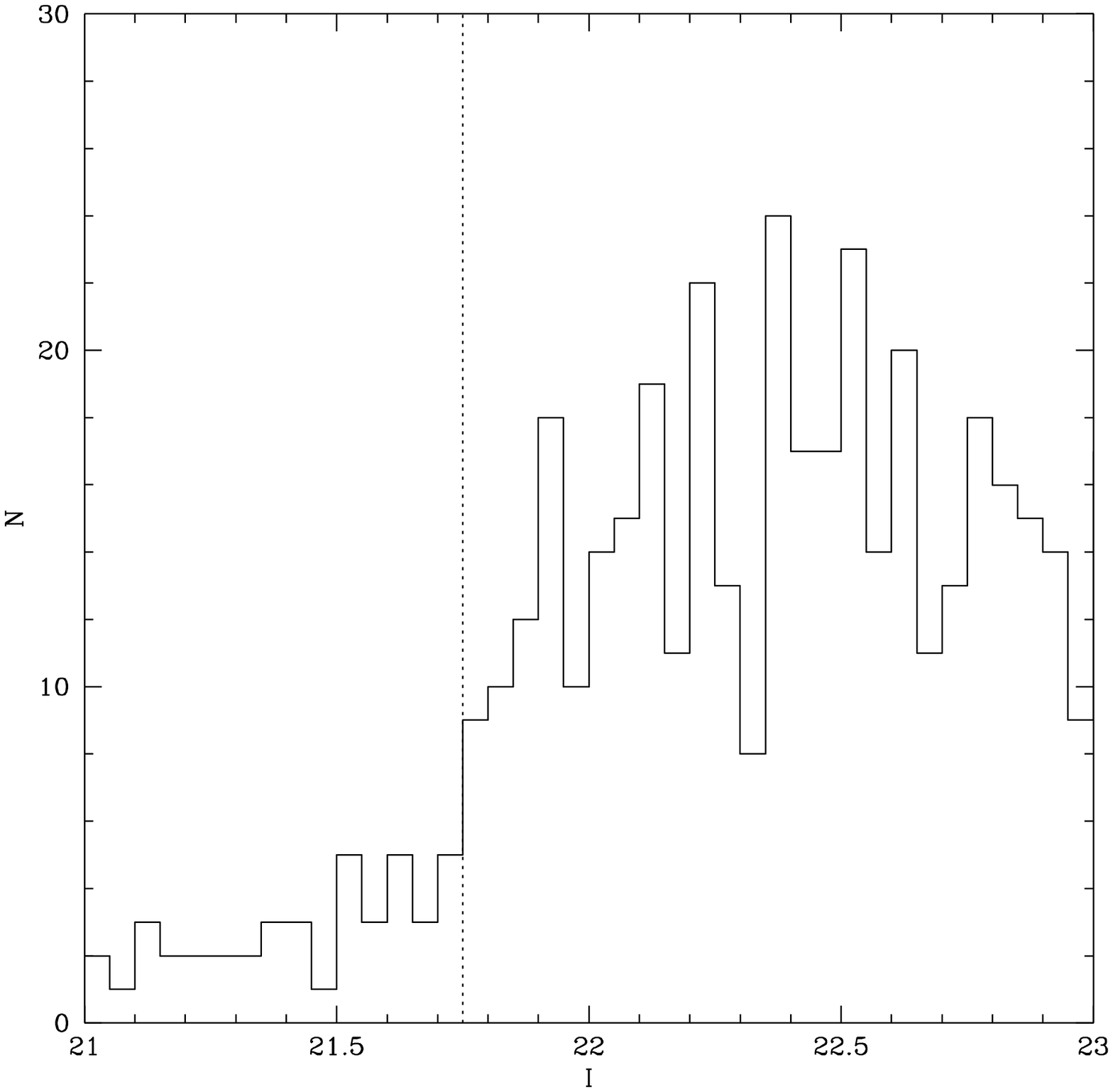}
\caption{$I$ luminosity function of the RGB tip.  To minimize confusion from red supergiants, color limits of $1.2 < V-I < 1.6$ were used to select stars for this measurement.  We measure the RGB tip at $I = 21.76 \pm 0.05$, as indicated by the dotted line. \label{figTRGB}}
\end{figure}

Our data contain two other distance indicators, the RGB tip and red clump, which can be used to provide independent checks of our Cepheid distance.  Figure \ref{figTRGB} shows the $I$-band luminosity function in the region of the RGB tip.  We measure the tip position to fall at $I = 21.76 \pm 0.05$, which is consistent with the measurements of $21.73 \pm 0.09$ by \citet{sak96} and $21.8 \pm 0.1$ by \citet{doh97}.  Correcting for extinction and adopting $M_I = -4.00 \pm 0.1$ computed from the Z=0.001 and Z=0.0004 Padova models \citep{gir00}, we calculate the true distance modulus to be $\mu_0 = 25.69 \pm 0.12$.

\begin{figure}
\plotone{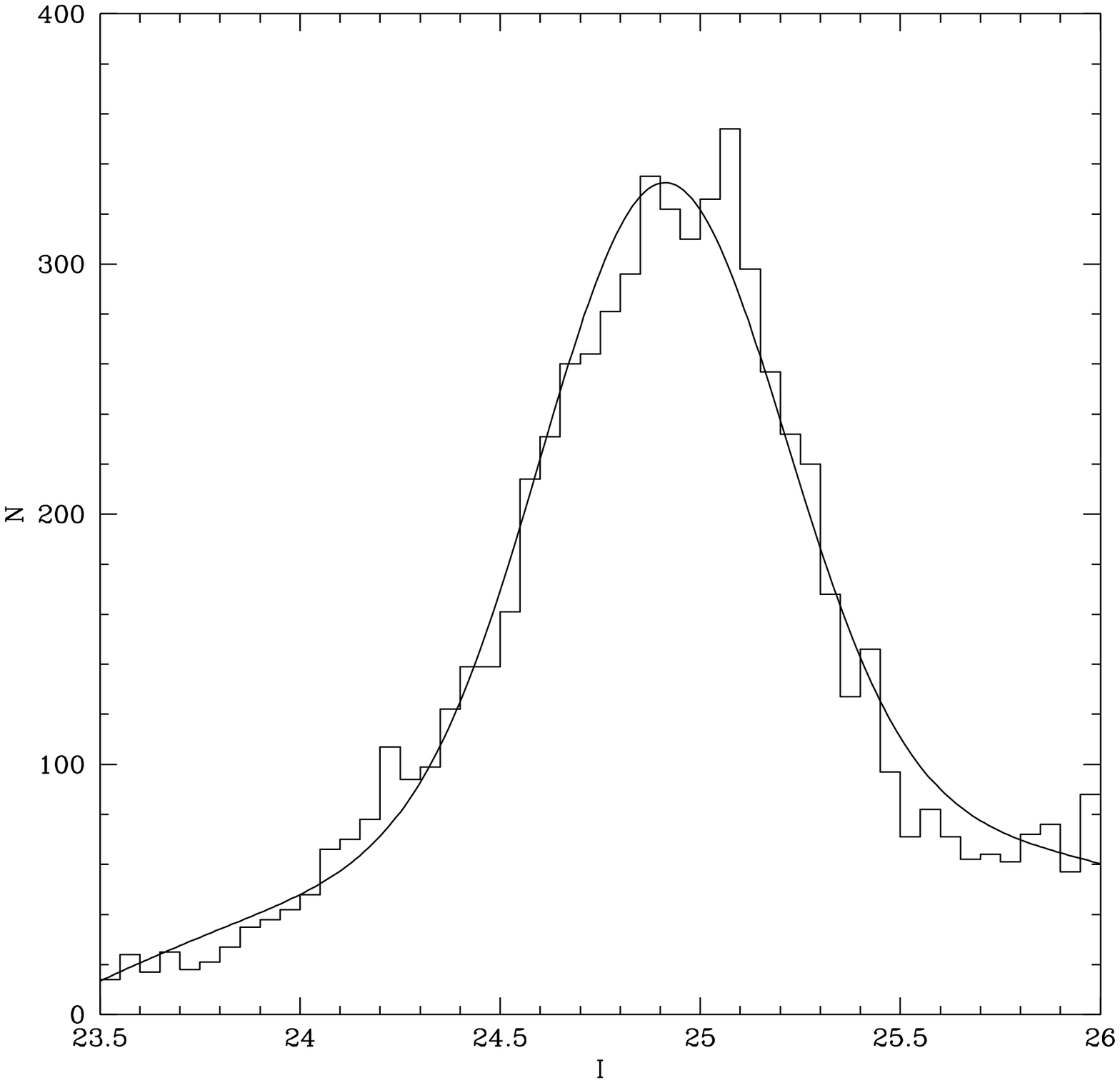}
\caption{$I$ luminosity function of the red clump.  Color cuts were $0.65 < V-I < 1.0$.  The curve indicates our best fit to the luminosity function, using the sum of a quadratic (to fit the underlying RGB stars) and a Gaussian (to fit the red clump).  The Gaussian is centered at $I = 24.91 \pm 0.01$, with a $1 \sigma$ width of $0.30 \pm 0.02$. \label{figRC}}
\end{figure}

Another independent distance can be determined from the position of the red clump.  Figure \ref{figRC} shows the $I$-band luminosity function in the region of the red clump.  Our best fit to the luminosity function was obtained with a Gaussian centered at $I = 24.91 \pm 0.01$ with a width of $\sigma_I = 0.30 \pm 0.02$.  Using the semi-empirical red clump calibration described by \citet{dol01a}, we calculate an absolute magnitude of $M_I = -0.67 \pm 0.15$ for Sextans A; this produces a true distance modulus of $\mu_0 = 25.51 \pm 0.15$.

Combining our three distance determinations, we calculate the true distance modulus to Sextans A to be $\mu_0 = 25.61 \pm 0.07$, corresponding to a distance of $d = 1.32 \pm 0.04$ Mpc.  We note that this distance is significantly closer than the value of 1.44 Mpc assumed in Paper I (adopted from Dohm-Palmer et al. 1997).  We also note that, while we do not explicitly include calibration uncertainties, these are $\sim 0.02$ magnitudes and thus do not add to our reported uncertainties.

Our distance is consistent with the \citet{pio94} Cepheid distance of ($\mu_0 = 25.71 \pm 0.20$), but is inconsistent at the $1 \sigma$ level with the \citet{sak96} Cepheid distance ($\mu_0 = 25.85 \pm 0.15$).  We note, however, that the \citet{sak96} Cepheid distance is based on single-epoch photometry and thus is more uncertain than is indicated by their error bars.  In addition, \citet{sak96} present an RGB tip distance modulus of $\mu_0 = 25.74 \pm 0.13$, which is consistent with our distance at the $1 \sigma$ level.

\section{Discussion}

\subsection{Short-Period Cepheids as Distance Indicators}

Distance measurement using Cepheids typically rely on longer-period Cepheids with ($P > 10$ days).  As we have observed in Sextans A and previously in Leo A, large numbers of short-period Cepheids are present in low-metallicity star-forming galaxies, due to the passage of the BHeB sequence through the instability strip at fainter magnitudes \citep{dol02}.  This allows for the determination of the distance using 82 Cepheids, which is potentially much more accurate than that using the 5 longer-period Cepheids of \citet{san82}.  However, one must first examine the question of whether or not these objects serve as reliable standard candles.

To address this question, we present distance moduli to five nearby galaxies in Table \ref{tabcompare} as determined by a variety of distance indicators.  Because of its relative insensitivity to age and metallicity, we will adopt the RGB tip distance as the comparison standard.  Comparing the short-period Cepheid distances with the RGB tip distances, we see no significant systematic differences for the low-metallicity galaxies (all but the LMC).  We do note that the first-overtone distances are shorter by an average of 0.08 magnitudes for the lowest-metallicity galaxies (Sextans A and Leo A).  This may be the sign of a real effect, but we also note that the data are consistent with no metallicity dependence of the distance.  At the $1 \sigma$ level, we can only conclude that any dependence of the short-period Cepheid P-L relations on metallicity is no more than $0.1$ magnitudes per dex.
\placetable{tabcompare}

However, the LMC distance from short-period Cepheids is inconsistent at the $2 \sigma$ level with that from the RGB tip (as well as from all other distance indicators), in that they are roughly $0.2$ magnitudes closer.  We interpret this as resulting from the metallicity dependence of the position of the BHeB sequence as mentioned above.  At sufficiently low metallicities, the heavily-populated BHeB sequence passes through the instability strip at a luminosity that produces large numbers of Cepheids with periods of $\sim 1$ day.  At higher metallicity, the BHeB sequence passes through the instability strip at a brighter magnitude, and thus 1-day Cepheids are only produced by stars passing through the instability strip as they evolve off the main sequence.  Based on these data, it appears that the dividing line is between the metallicities of the SMC and LMC.  The EROS P-L relations \citep{bau99} give strength to the argument that there is a fundamental difference between the LMC and SMC Cepheid populations, as they found a significant slope change at 2 days in their SMC Cepheids but not their LMC ones.

\subsection{Comparisons of Other Distance Indicators}

Table \ref{tabcompare} also gives the distances as calculated using RR Lyraes, long-period Cepheids, and the red clump.  This allows us to expand on the comparison we made in an earlier paper \citep{dol01a}, which considered only IC 1613 and the Magellanic Clouds.

The four RR Lyrae samples are all taken from objects with $-1.9 < \logz < -1.3$, thus providing an insufficient baseline for an examination of the effects of metallicity on the RR Lyrae calibration.  However, we can compare the zero point of the calibration to that of the RGB tip, and we find that the values we assumed are consistent at much better than the $1 \sigma$ level.

A comparison of the four long-period Cepheid distances with the RGB tip distances shows a hint of a metallicity dependence in the Cepheid distances.  Making a least $\chi^2$ fit, we measure a slope of $-0.12 \pm 0.12$ magnitudes per dex, in the sense that low-metallicity galaxies are measured to be more distant by Cepheids than by the RGB tip.  This slope is consistent with those predicted theoretically by \citet{san99} and measured by \citet{ken98}.  While this is a $1 \sigma$ result, improved data on the Sextans A long-period Cepheids (or those in similarly metal-poor galaxies) is needed to improve the measurement.  It is clear that the P-L relation does not have a large ($\sim 0.5$ magnitudes per dex; Beaulieu et al. 1997 and Gould 1994) dependence on metallicity.

We find no evidence of a correlation between metallicity and the red clump $-$ RGB tip distance difference, which is comforting since metallicity is accounted for in our semi-empirical red clump calibration.  The zero points are also consistent at the $1 \sigma$ level, providing additional evidence that the red clump is an accurate distance indicator, provided that population effects are properly accounted for.

Finally, we present distance moduli relative to the SMC for all galaxies in Table \ref{tabrelative}.  Although the LMC is generally the standard for such comparisons, we chose to use the SMC for the comparison because it has accurate distances determined with all five distance indicators.
\placetable{tabrelative}

\section{Summary}

We have presented photometry of deep WFPC2 observations of Sextans A.  Observations were made over 16 epochs in F814W and 8 epochs in F555W, permitting a search for variable stars.  We identified 82 periodic variables with clean light curves in both bands; all 82 appear to be short-period Cepheids.  We found fundamental-mode Cepheids with periods as short as 0.8 days and first-overtone Cepheids with periods of 0.5 days.

Using these Cepheids, we measured a distance to Sextans A using a P-L relation computed using SMC short-period ($\sim 1$ day) Cepheids.  We compared this distance to that obtained with other means, and find that short-period Cepheids are indeed a viable standard candle for objects at or below the metallicity of the SMC.  Given the large numbers of these that should be found in low-metallicity star-forming galaxies, these objects can provide much more accurate distance measurements than those obtained using other distance indicators.

Combining the short-period Cepheids with distances obtained from the RGB tip and red clump, we determined the distance modulus to Sextans A to be $\mu_0 = 25.61 \pm 0.07$, corresponding to a distance of $d = 1.32 \pm 0.04$ Mpc.

We also examined relative distances produced by five distance indicators for five galaxies in and near the Local Group: Sextans A, Leo A, IC 1613, and the Magellanic Clouds.  We find that relative distances calculated using the RGB tip, red clump, and RR Lyrae are consistent for the sample.  However, we find a metallicity dependence of $-0.12 \pm 0.12$ magnitudes per dex in distances calculated using reddening-free Cepheid techniques.  The relative distances for the five galaxies are calculated using these distance measurements.

\acknowledgments

Support for this work was provided by NASA through grant number GO-07496 from the Space Telescope Science Institute, which is operated by AURA, Inc., under NASA contract NAS 5-26555. EDS is grateful for partial support from NASA LTSARP grant No. NAG5-9221.

\clearpage
\begin{figure}
\caption{Deep image of Sextans A \label{figImage}}
\end{figure}

\begin{figure}
\caption{$(V-I)$, $I$ Color-magnitude diagram (33681 stars), calculated from mean magnitudes in all epochs. Poorly-fit stars ($\chi > 2$ or $|$sharpness$| > 0.3$) are not included.  Absolute magnitudes (on the right y-axis) are calculated assuming $I - M_I = 25.67$. \label{figCMD}}
\end{figure}

\def\fnum@figure{{\rmfamily Fig.\space\thefigure.---}}%
\clearpage
\begin{figure}
\plotone{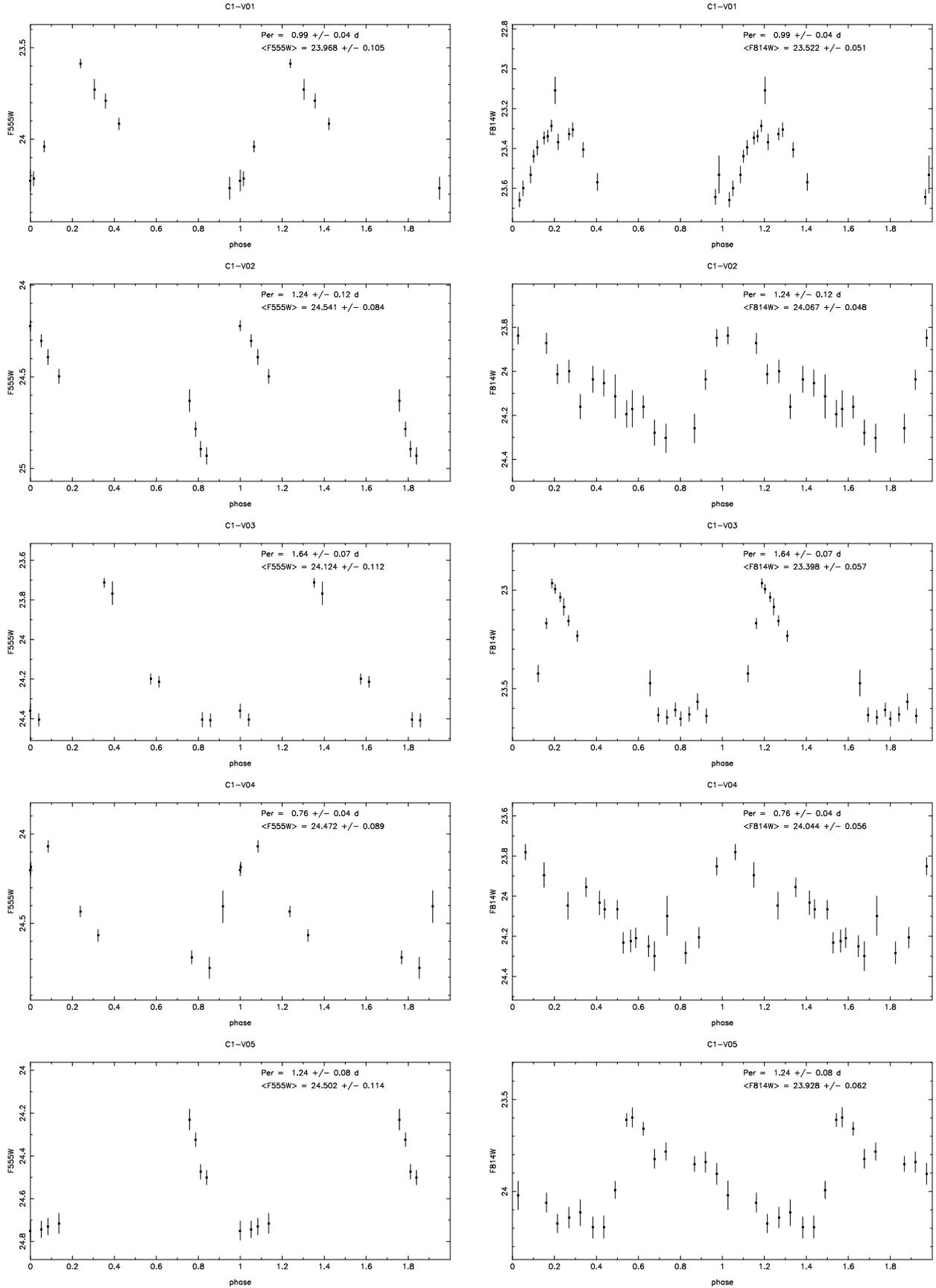}
\caption{Light curves of 82 variable stars in Sextans A. \label{figLCs}}
\end{figure}

\clearpage
\begin{figure}
\plotone{dolphin.fig3b.ps}
\addtocounter{figure}{-1}
\caption{Continued}
\end{figure}

\clearpage
\begin{figure}
\plotone{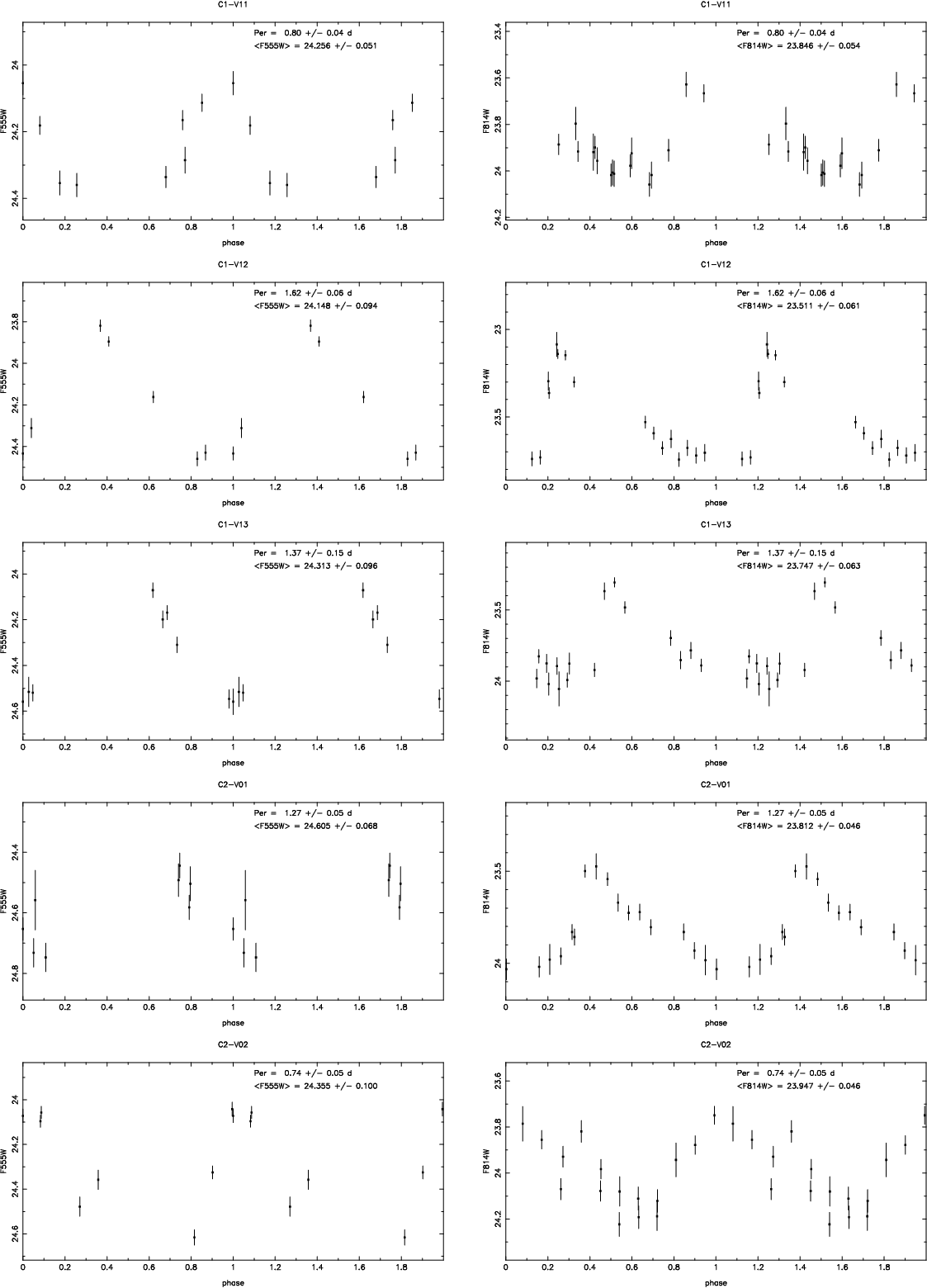}
\addtocounter{figure}{-1}
\caption{Continued}
\end{figure}

\clearpage
\begin{figure}
\plotone{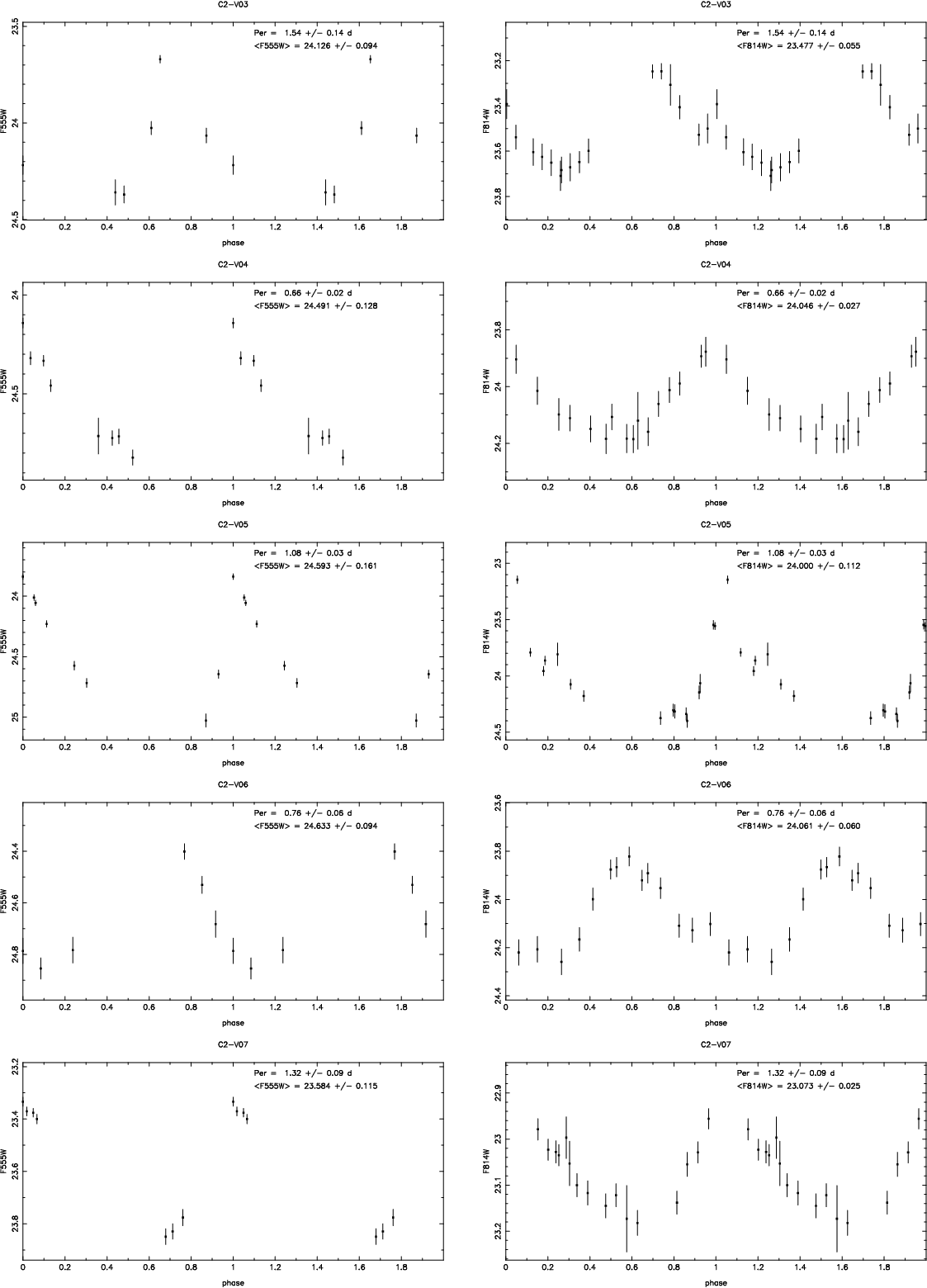}
\addtocounter{figure}{-1}
\caption{Continued}
\end{figure}

\clearpage
\begin{figure}
\plotone{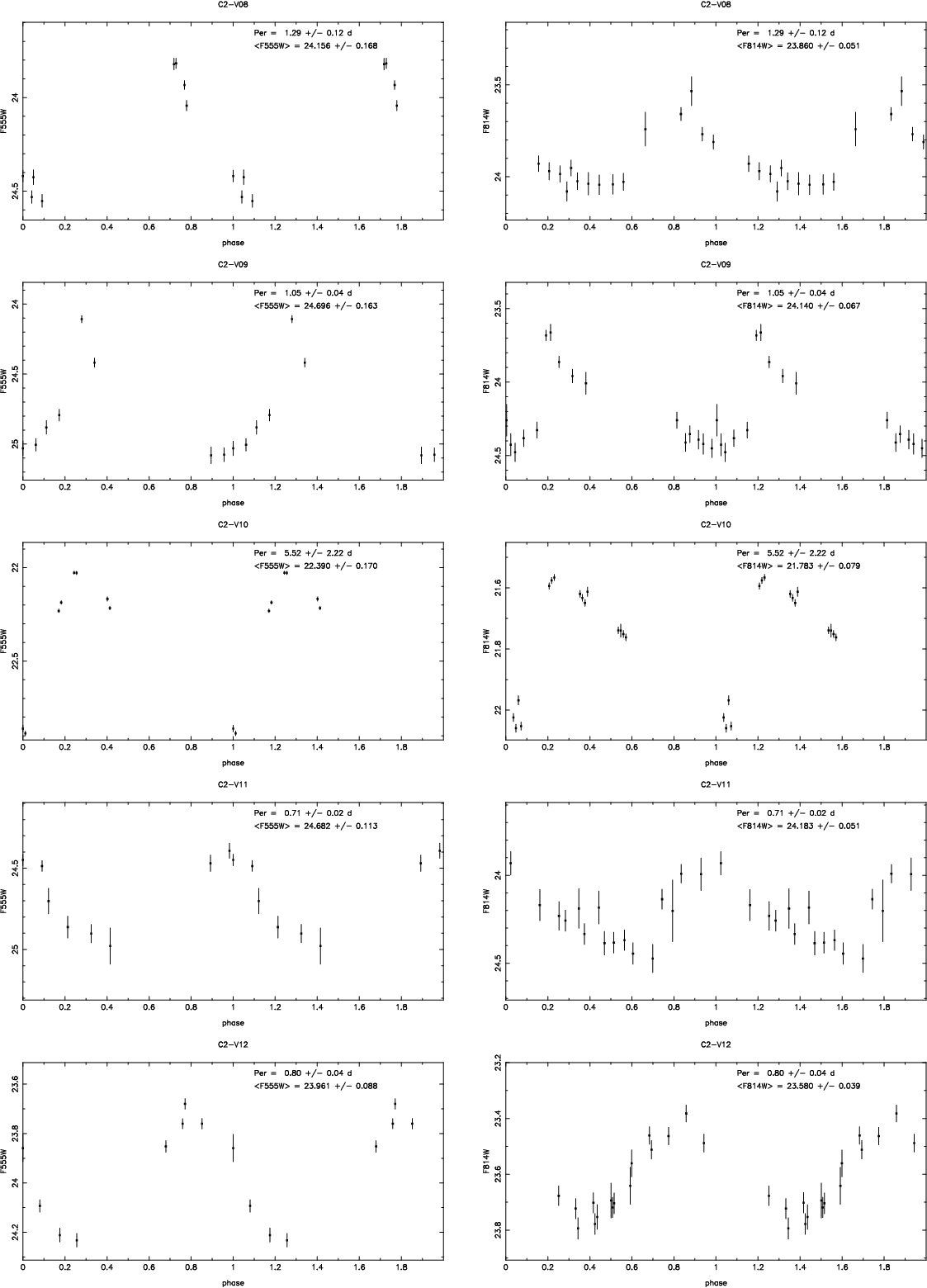}
\addtocounter{figure}{-1}
\caption{Continued}
\end{figure}

\clearpage
\begin{figure}
\plotone{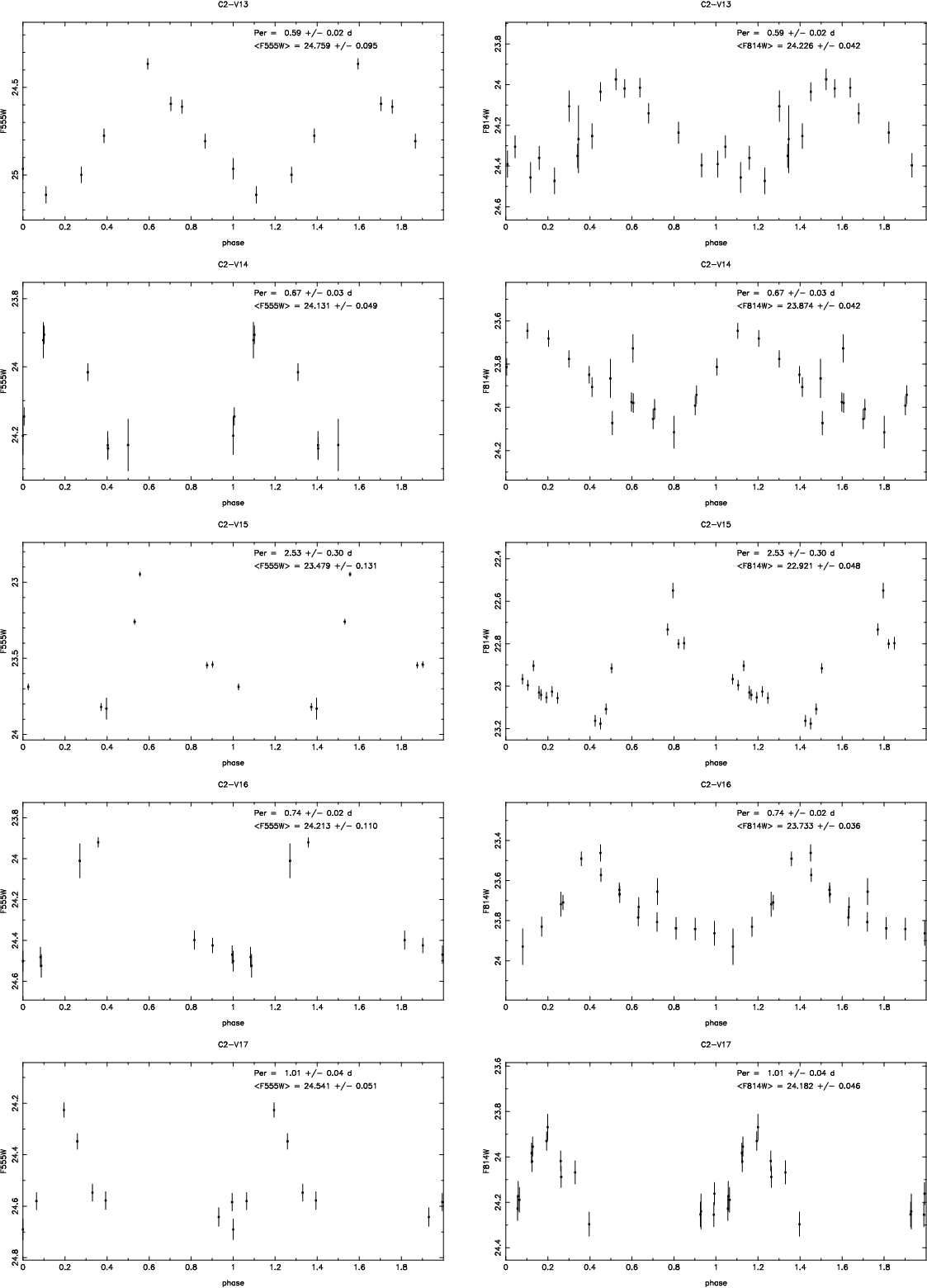}
\addtocounter{figure}{-1}
\caption{Continued}
\end{figure}

\clearpage
\begin{figure}
\plotone{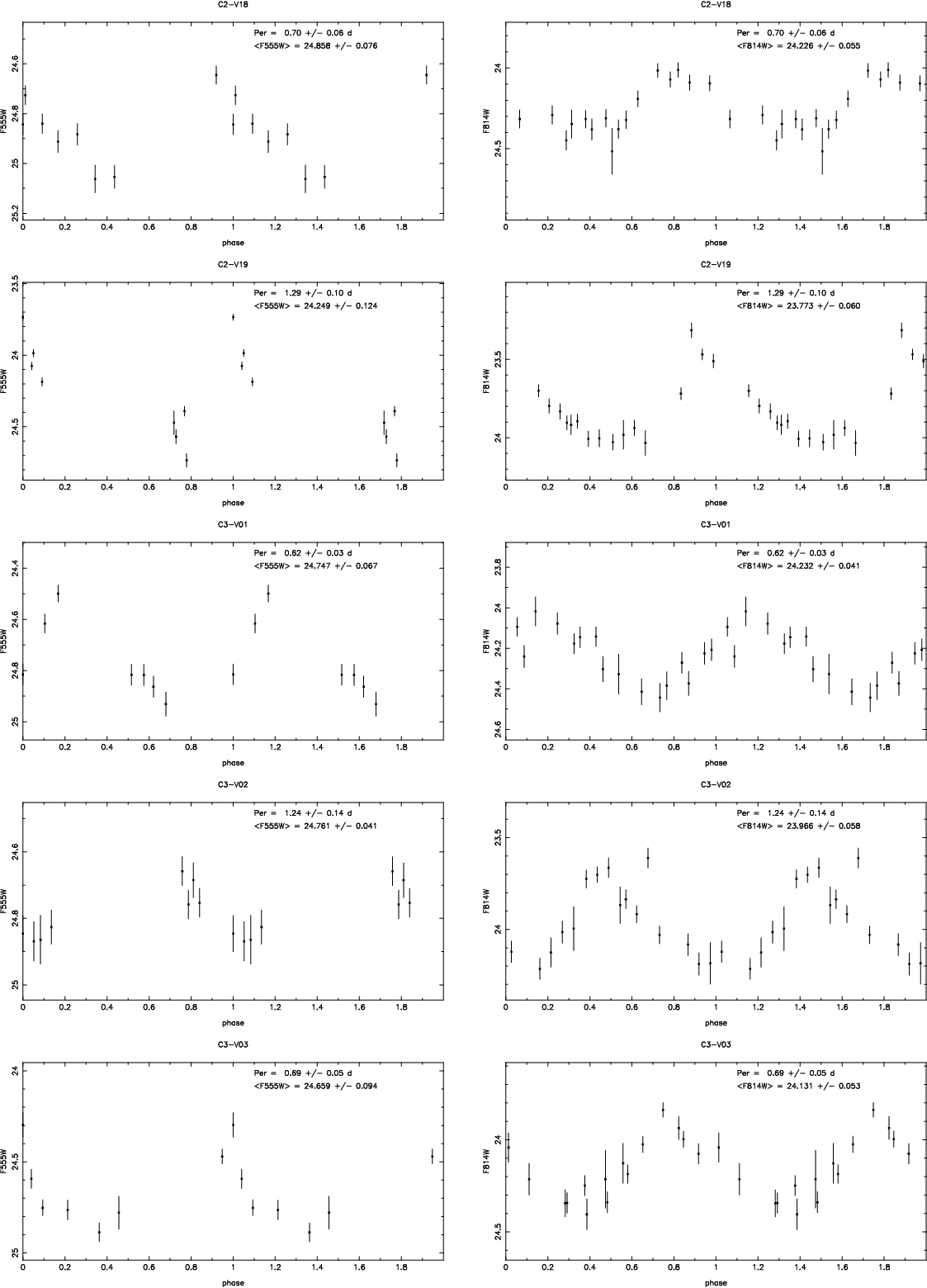}
\addtocounter{figure}{-1}
\caption{Continued}
\end{figure}

\clearpage
\begin{figure}
\plotone{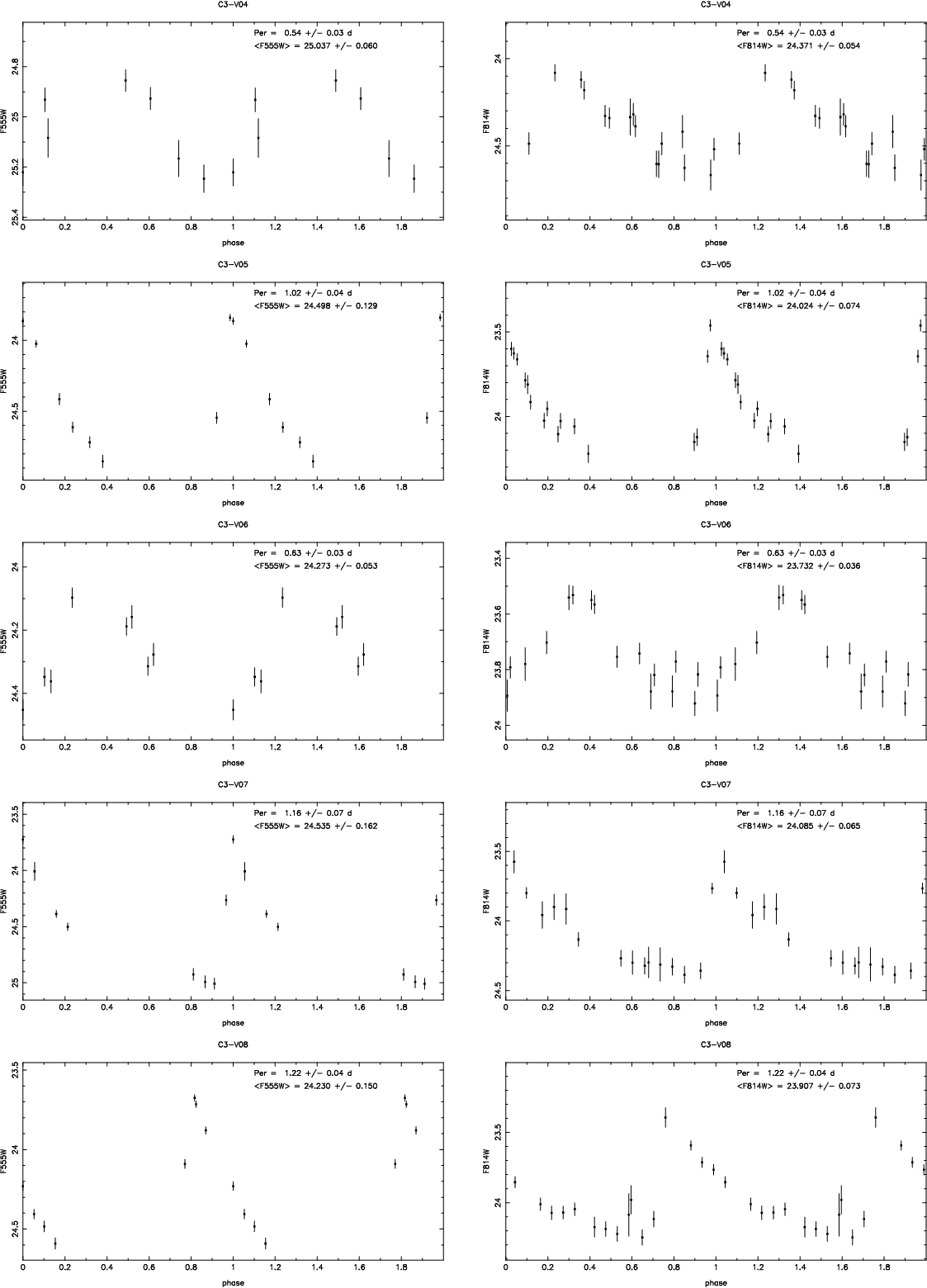}
\addtocounter{figure}{-1}
\caption{Continued}
\end{figure}

\clearpage
\begin{figure}
\plotone{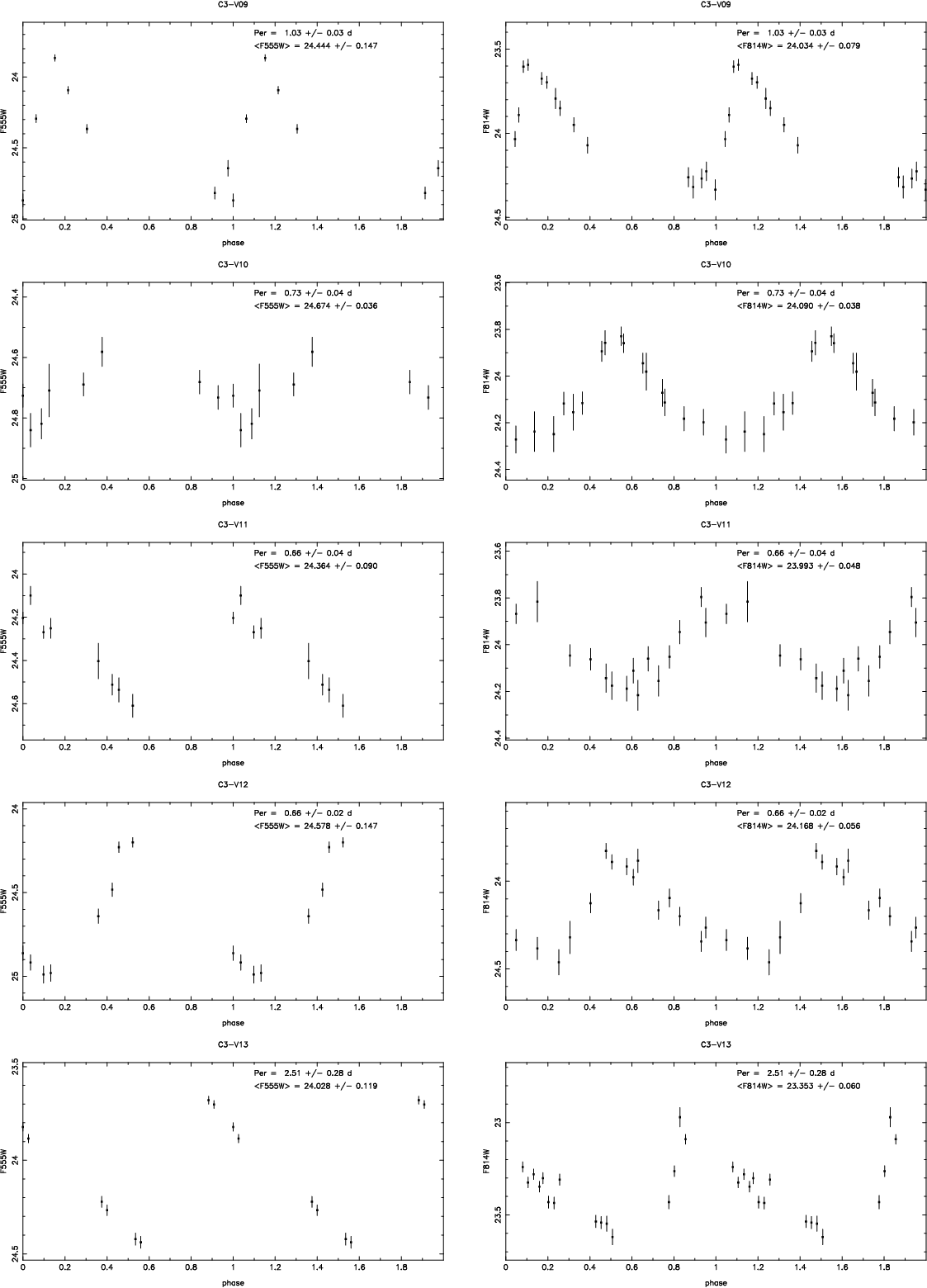}
\addtocounter{figure}{-1}
\caption{Continued}
\end{figure}

\clearpage
\begin{figure}
\plotone{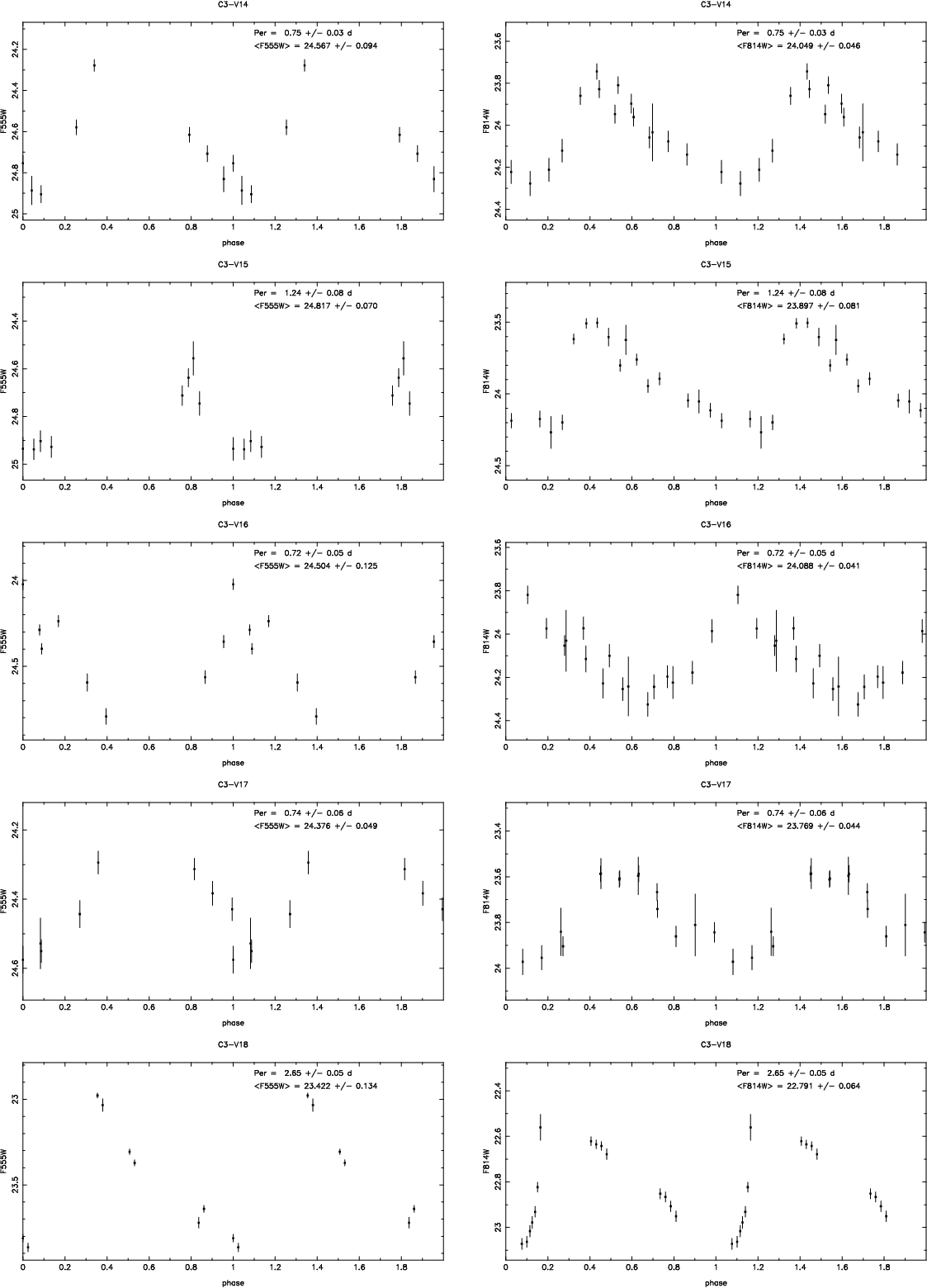}
\addtocounter{figure}{-1}
\caption{Continued}
\end{figure}

\clearpage
\begin{figure}
\plotone{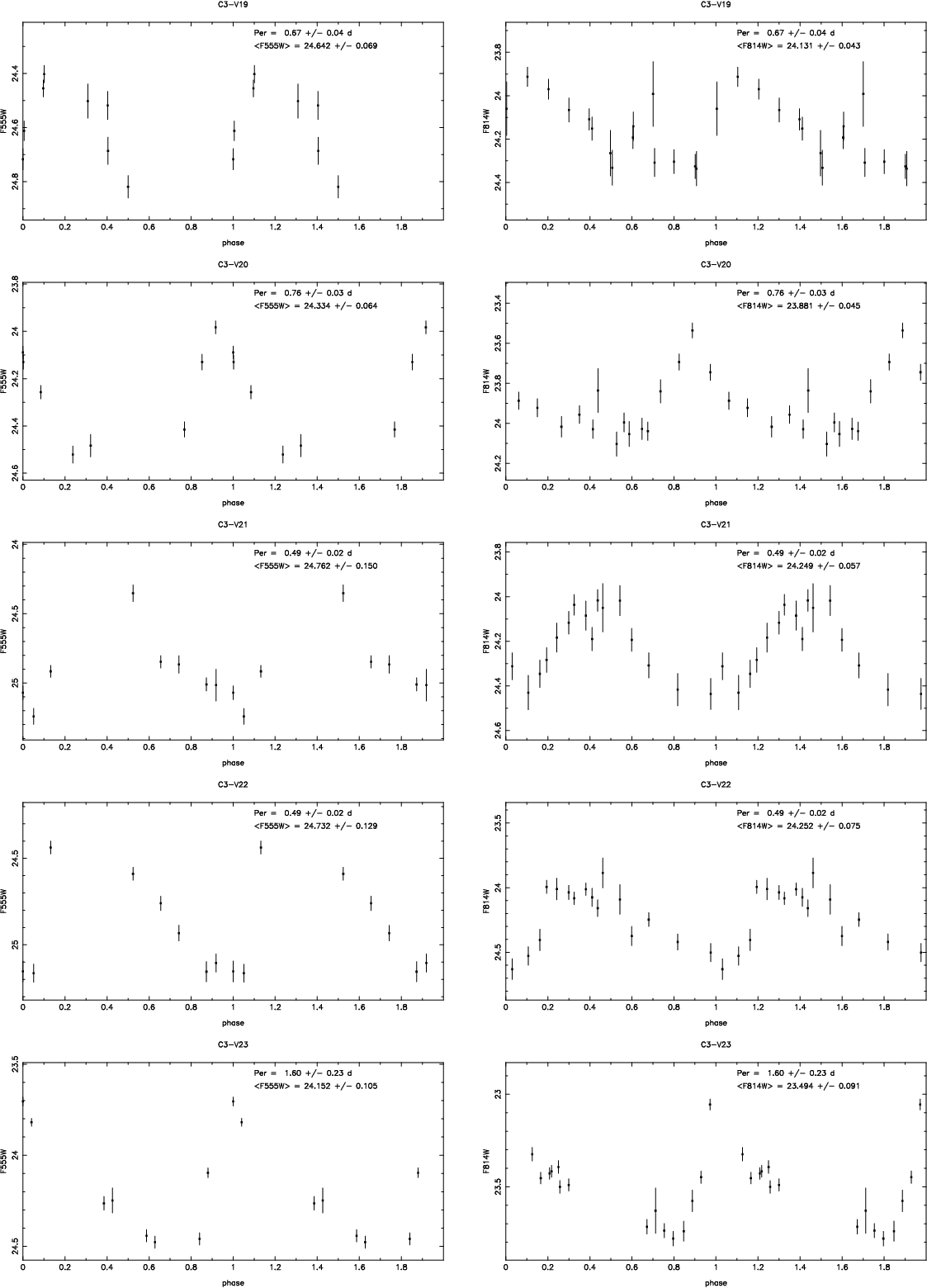}
\addtocounter{figure}{-1}
\caption{Continued}
\end{figure}

\clearpage
\begin{figure}
\plotone{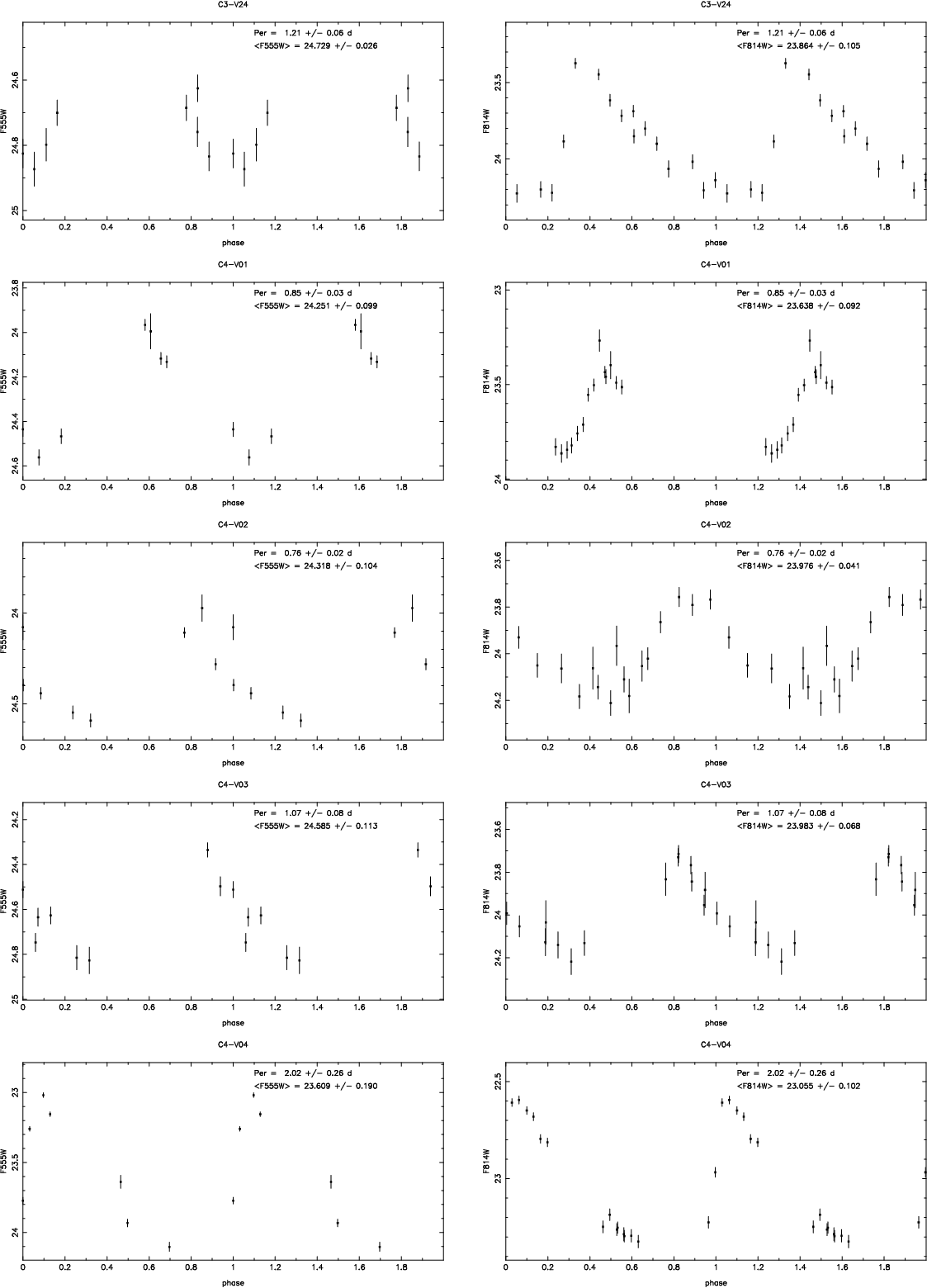}
\addtocounter{figure}{-1}
\caption{Continued}
\end{figure}

\clearpage
\begin{figure}
\plotone{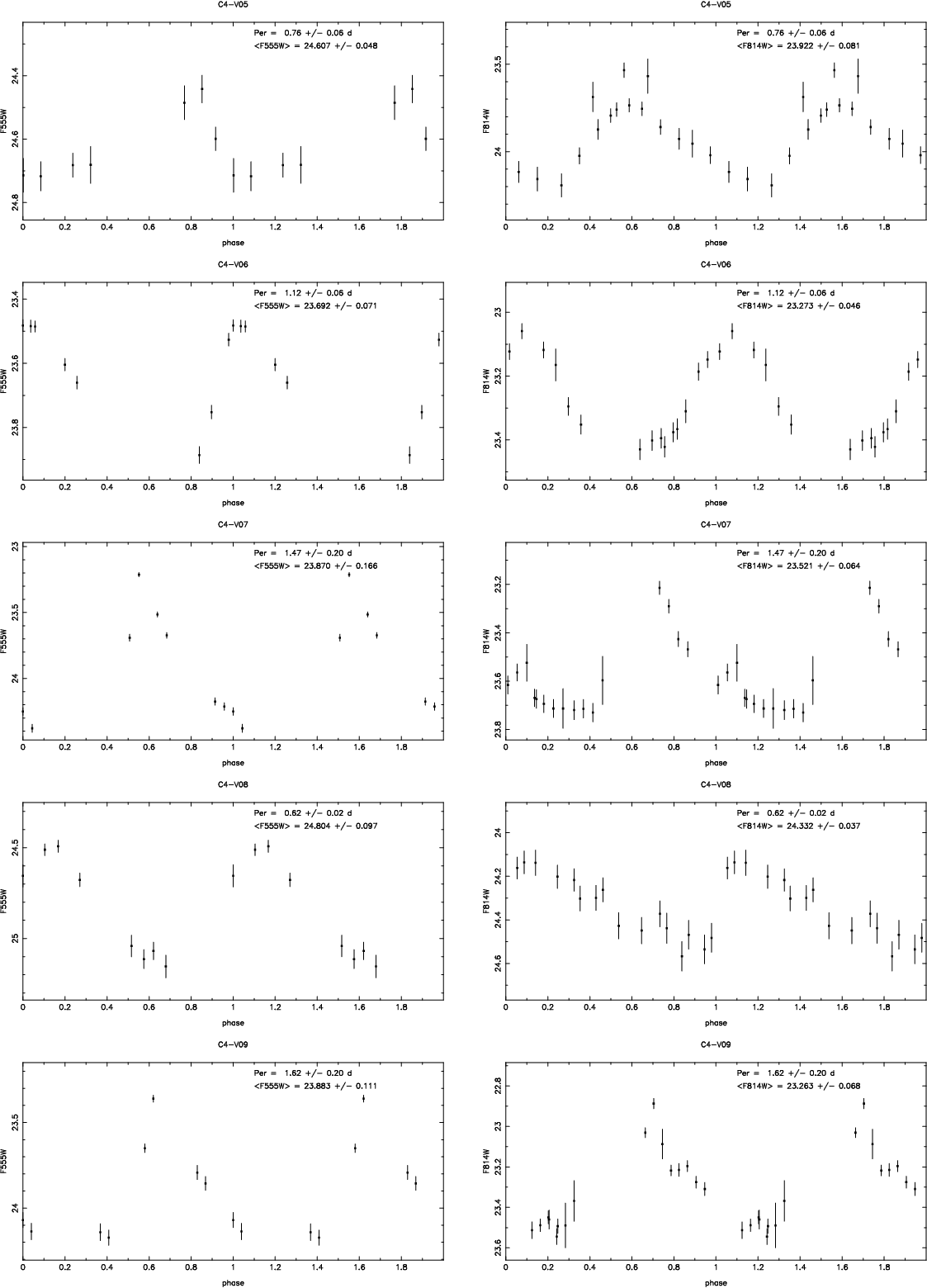}
\addtocounter{figure}{-1}
\caption{Continued}
\end{figure}

\clearpage
\begin{figure}
\plotone{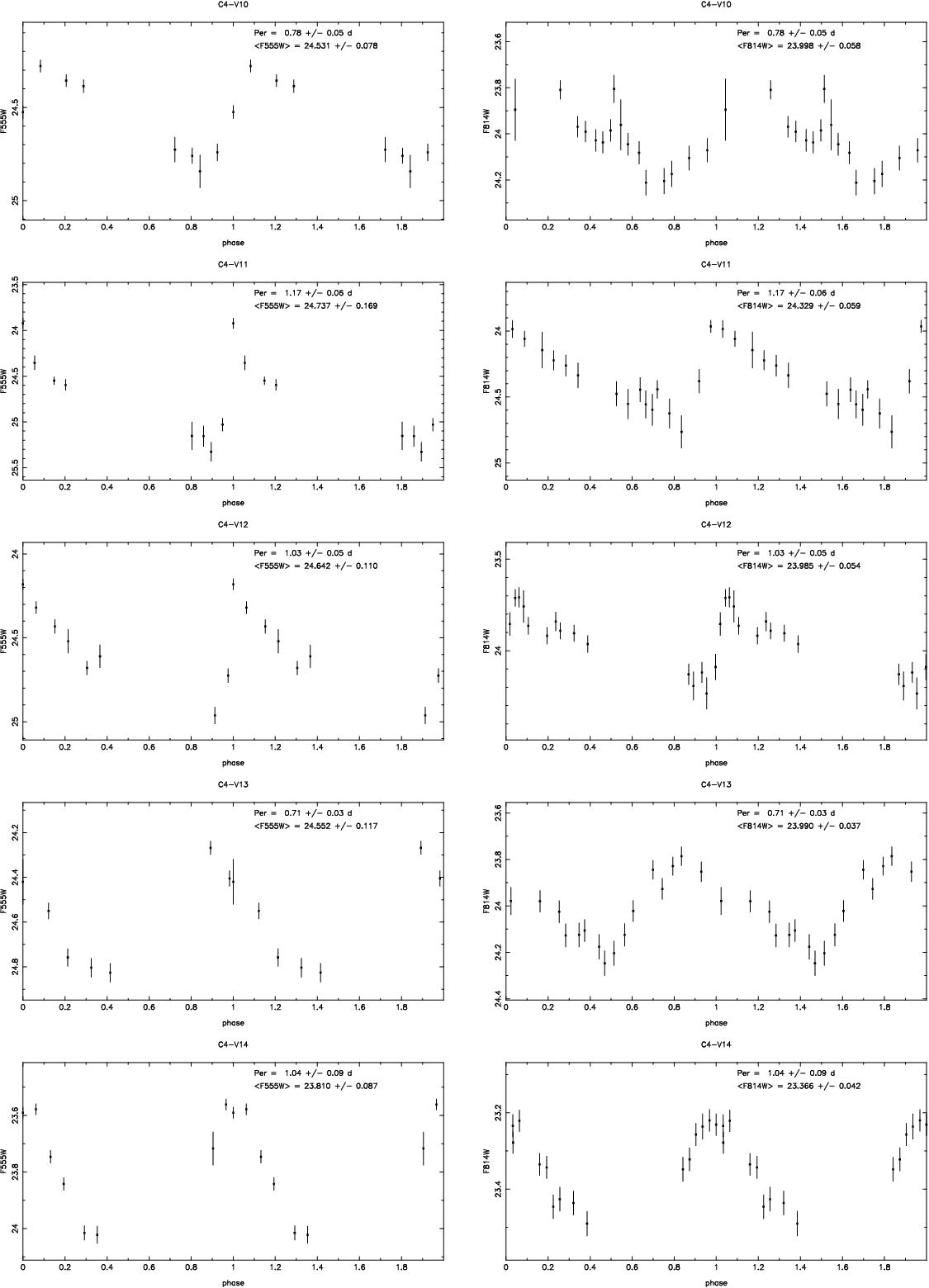}
\addtocounter{figure}{-1}
\caption{Continued}
\end{figure}

\clearpage
\begin{figure}
\plotone{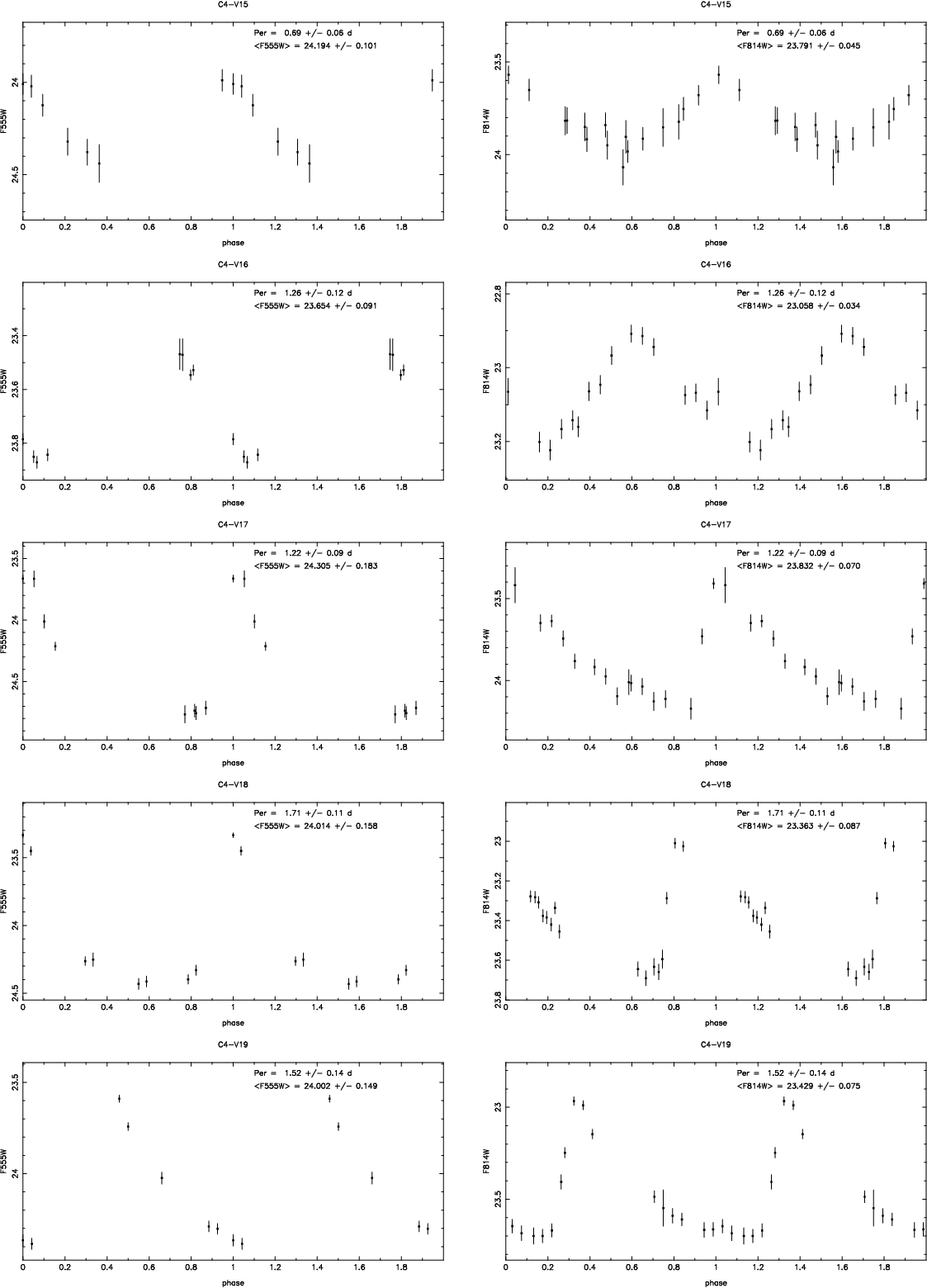}
\addtocounter{figure}{-1}
\caption{Continued}
\end{figure}

\clearpage
\begin{figure}
\plotone{dolphin.fig3p.ps}
\addtocounter{figure}{-1}
\caption{Continued}
\end{figure}

\clearpage
\begin{figure}
\plotone{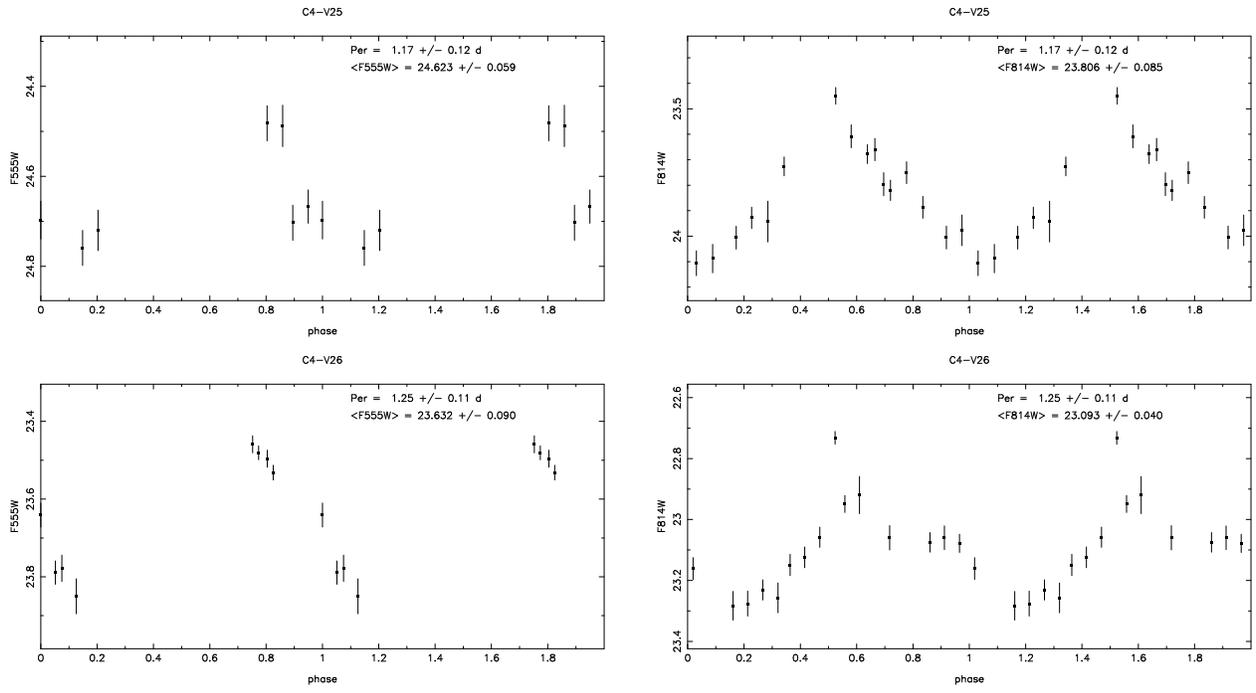}
\addtocounter{figure}{-1}
\caption{Continued}
\end{figure}

\clearpage
\begin{figure}
\caption{CMD of Sextans A field, with the detected variable stars shown as circles.  The detection efficiency was not 100\%, and thus there are many additional stars in the instability strip that are likely variable but were not classified as such.  The excess color spread of short period variables is artificially broad due to our limited phase coverage in $V$ (see text).  Absolute magnitudes (on the right y-axis) are calculated assuming $I - M_I = 25.67$. \label{figVCMD}}
\end{figure}

\clearpage

\clearpage

\clearpage

\clearpage

\clearpage
\begin{table}
\begin{center}
\addtocounter{table}{1}
Table \thetable.  Photometry of Variable Stars. \label{tab_phot}
\begin{tabular}{cccccc}
\\
\tableline
\tableline
         &         C1-V01 &         C1-V02 &         C1-V03 &         C1-V04 &         C1-V05 \\
\tableline
HJD$^a$  & F555W          & F555W          & F555W          & F555W          & F555W          \\
273.6254 & $24.23\pm0.06$ & $24.22\pm0.03$ & $24.36\pm0.04$ & $24.20\pm0.03$ & $24.75\pm0.04$ \\
273.6900 & $24.04\pm0.03$ & $24.30\pm0.03$ & $24.41\pm0.03$ & $24.07\pm0.03$ & $24.74\pm0.04$ \\
274.5657 & $24.27\pm0.06$ & $24.63\pm0.06$ & $24.20\pm0.03$ & $24.43\pm0.03$ & $24.23\pm0.05$ \\
274.6302 & $24.22\pm0.04$ & $24.89\pm0.04$ & $24.21\pm0.03$ & $24.57\pm0.03$ & $24.47\pm0.03$ \\
274.9691 & $23.79\pm0.04$ & $24.39\pm0.04$ & $24.41\pm0.04$ & $24.69\pm0.04$ & $24.73\pm0.04$ \\
\end{tabular}
\end{center}
$^a$ HJD$-$2451000
\end{table}
\clearpage
\begin{table}
\begin{center}
Table \thetable---Continued \\
\begin{tabular}{cccccc}
\\
\tableline
\tableline
         &         C1-V06 &         C1-V07 &         C1-V08 &         C1-V09 &         C1-V10 \\
\tableline
HJD$^a$  & F555W          & F555W          & F555W          & F555W          & F555W          \\
273.6254 & $24.66\pm0.04$ & $24.91\pm0.04$ & $25.17\pm0.05$ & $23.30\pm0.02$ & $25.15\pm0.06$ \\
273.6900 & $24.76\pm0.04$ & $24.80\pm0.04$ & $25.01\pm0.05$ & $23.25\pm0.02$ & $25.14\pm0.05$ \\
274.5657 & $24.98\pm0.04$ & $24.98\pm0.05$ & $25.27\pm0.07$ & $23.65\pm0.06$ & $24.61\pm0.05$ \\
274.6302 & $24.99\pm0.05$ & $24.94\pm0.04$ & $25.23\pm0.06$ & $23.64\pm0.02$ & $24.77\pm0.04$ \\
274.9691 & $24.92\pm0.05$ & $24.59\pm0.04$ & $24.19\pm0.03$ & $23.61\pm0.02$ & $25.15\pm0.06$ \\
\end{tabular}
\end{center}
\end{table}
\clearpage
\begin{table}
\begin{center}
Table \thetable---Continued \\
\begin{tabular}{cccccc}
\\
\tableline
\tableline
         &         C1-V11 &         C1-V12 &         C1-V13 &         C2-V01 &         C2-V02 \\
\tableline
HJD$^a$  & F555W          & F555W          & F555W          & F555W          & F555W          \\
273.6254 & $24.05\pm0.04$ & $24.43\pm0.03$ & $24.56\pm0.06$ & $24.65\pm0.04$ & $24.07\pm0.03$ \\
273.6900 & $24.18\pm0.03$ & $24.31\pm0.05$ & $24.52\pm0.04$ & $24.73\pm0.05$ & $24.06\pm0.03$ \\
274.5657 & $24.35\pm0.04$ &     . . .      & $24.17\pm0.03$ & $24.49\pm0.05$ & $24.48\pm0.04$ \\
274.6302 & $24.36\pm0.04$ & $24.16\pm0.03$ & $24.31\pm0.03$ & $24.58\pm0.04$ & $24.36\pm0.04$ \\
274.9691 & $24.34\pm0.03$ & $24.46\pm0.03$ & $24.55\pm0.04$ & $24.56\pm0.10$ & $24.62\pm0.03$ \\
\end{tabular}
\end{center}
\end{table}
\clearpage
\begin{table}
\begin{center}
Table \thetable---Continued \\
\begin{tabular}{cccccc}
\\
\tableline
\tableline
         &         C2-V03 &         C2-V04 &         C2-V05 &         C2-V06 &         C2-V07 \\
\tableline
HJD$^a$  & F555W          & F555W          & F555W          & F555W          & F555W          \\
273.6254 & $24.22\pm0.05$ & $24.14\pm0.03$ & $23.84\pm0.02$ & $24.79\pm0.05$ & $23.33\pm0.02$ \\
273.6900 &     . . .      & $24.33\pm0.03$ & $24.05\pm0.02$ & $24.85\pm0.04$ & $23.38\pm0.02$ \\
274.5657 & $24.03\pm0.03$ & $24.72\pm0.04$ & $25.03\pm0.05$ & $24.78\pm0.05$ & $23.83\pm0.03$ \\
274.6302 & $23.67\pm0.02$ & $24.82\pm0.04$ & $24.64\pm0.03$ &     . . .      & $23.78\pm0.03$ \\
274.9691 & $24.07\pm0.04$ & $24.32\pm0.03$ & $24.57\pm0.04$ & $24.40\pm0.03$ & $23.37\pm0.02$ \\
\end{tabular}
\end{center}
\end{table}
\clearpage
\begin{table}
\begin{center}
Table \thetable---Continued \\
\begin{tabular}{cccccc}
\\
\tableline
\tableline
         &         C2-V08 &         C2-V09 &         C2-V10 &         C2-V11 &         C2-V12 \\
\tableline
HJD$^a$  & F555W          & F555W          & F555W          & F555W          & F555W          \\
273.6254 & $24.42\pm0.03$ & $25.03\pm0.05$ & $22.86\pm0.02$ & $24.45\pm0.04$ & $23.86\pm0.06$ \\
273.6900 & $24.43\pm0.04$ & $25.00\pm0.04$ & $22.89\pm0.01$ & $24.49\pm0.03$ & $24.09\pm0.03$ \\
274.5657 & $23.82\pm0.03$ & $25.08\pm0.06$ & $22.23\pm0.01$ & $24.90\pm0.06$ & $24.21\pm0.03$ \\
274.6302 & $24.04\pm0.03$ & $25.08\pm0.05$ & $22.19\pm0.01$ & $24.98\pm0.11$ & $24.23\pm0.03$ \\
274.9691 & $24.53\pm0.03$ & $24.11\pm0.03$ & $22.03\pm0.01$ & $24.47\pm0.05$ & $23.85\pm0.02$ \\
\end{tabular}
\end{center}
\end{table}
\clearpage
\begin{table}
\begin{center}
Table \thetable---Continued \\
\begin{tabular}{cccccc}
\\
\tableline
\tableline
         &         C2-V13 &         C2-V14 &         C2-V15 &         C2-V16 &         C2-V17 \\
\tableline
HJD$^a$  & F555W          & F555W          & F555W          & F555W          & F555W          \\
273.6254 & $24.96\pm0.06$ & $24.20\pm0.06$ &     . . .      & $24.50\pm0.05$ & $24.69\pm0.04$ \\
273.6900 & $25.11\pm0.05$ & $23.92\pm0.05$ & $23.69\pm0.02$ & $24.52\pm0.06$ & $24.58\pm0.03$ \\
274.5657 & $24.37\pm0.03$ & $24.23\pm0.04$ & $23.82\pm0.02$ & $24.01\pm0.08$ & $24.64\pm0.04$ \\
274.6302 & $24.59\pm0.04$ & $24.23\pm0.08$ & $23.83\pm0.07$ & $23.92\pm0.02$ & $24.58\pm0.03$ \\
274.9691 & $25.00\pm0.04$ & $24.15\pm0.03$ & $23.26\pm0.02$ & $24.40\pm0.04$ & $24.55\pm0.03$ \\
\end{tabular}
\end{center}
\end{table}
\clearpage
\begin{table}
\begin{center}
Table \thetable---Continued \\
\begin{tabular}{cccccc}
\\
\tableline
\tableline
         &         C2-V18 &         C2-V19 &         C3-V01 &         C3-V02 &         C3-V03 \\
\tableline
HJD$^a$  & F555W          & F555W          & F555W          & F555W          & F555W          \\
273.6254 & $24.84\pm0.04$ & $23.73\pm0.02$ & $24.82\pm0.04$ & $24.85\pm0.05$ & $24.30\pm0.07$ \\
273.6900 & $24.84\pm0.04$ & $23.98\pm0.02$ & $24.62\pm0.04$ & $24.87\pm0.06$ & $24.75\pm0.04$ \\
274.5657 & $25.06\pm0.06$ & $24.57\pm0.05$ & $24.82\pm0.04$ & $24.66\pm0.04$ & $24.89\pm0.05$ \\
274.6302 & $25.05\pm0.05$ & $24.73\pm0.05$ & $24.86\pm0.04$ & $24.68\pm0.05$ & $24.78\pm0.09$ \\
274.9691 & $24.64\pm0.04$ & $24.07\pm0.03$ & $24.50\pm0.03$ & $24.86\pm0.07$ & $24.47\pm0.04$ \\
\end{tabular}
\end{center}
\end{table}
\clearpage
\begin{table}
\begin{center}
Table \thetable---Continued \\
\begin{tabular}{cccccc}
\\
\tableline
\tableline
         &         C3-V04 &         C3-V05 &         C3-V06 &         C3-V07 &         C3-V08 \\
\tableline
HJD$^a$  & F555W          & F555W          & F555W          & F555W          & F555W          \\
273.6254 & $25.22\pm0.05$ & $23.86\pm0.02$ & $24.45\pm0.03$ & $23.72\pm0.03$ & $24.23\pm0.03$ \\
273.6900 & $25.08\pm0.08$ & $24.02\pm0.02$ & $24.35\pm0.03$ & $24.01\pm0.08$ & $24.41\pm0.03$ \\
274.5657 & $25.17\pm0.07$ & $24.55\pm0.04$ & $24.19\pm0.03$ & $24.93\pm0.05$ & $24.09\pm0.03$ \\
274.6302 & $25.25\pm0.06$ & $23.84\pm0.02$ & $24.31\pm0.03$ & $24.99\pm0.05$ & $23.71\pm0.02$ \\
274.9691 & $24.86\pm0.04$ & $24.72\pm0.04$ & $24.36\pm0.04$ & $24.39\pm0.03$ & $24.48\pm0.03$ \\
\end{tabular}
\end{center}
\end{table}
\clearpage
\begin{table}
\begin{center}
Table \thetable---Continued \\
\begin{tabular}{cccccc}
\\
\tableline
\tableline
         &         C3-V09 &         C3-V10 &         C3-V11 &         C3-V12 &         C3-V13 \\
\tableline
HJD$^a$  & F555W          & F555W          & F555W          & F555W          & F555W          \\
273.6254 & $24.87\pm0.05$ & $24.73\pm0.04$ & $24.20\pm0.03$ & $24.86\pm0.04$ & $23.82\pm0.02$ \\
273.6900 & $24.29\pm0.03$ & $24.82\pm0.05$ & $24.27\pm0.03$ & $24.99\pm0.05$ & $23.88\pm0.02$ \\
274.5657 & $24.82\pm0.04$ & $24.69\pm0.04$ & $24.51\pm0.05$ & $24.48\pm0.04$ & $24.22\pm0.03$ \\
274.6302 & $24.64\pm0.06$ & $24.58\pm0.05$ & $24.61\pm0.05$ & $24.20\pm0.03$ & $24.27\pm0.03$ \\
274.9691 & $24.37\pm0.03$ & $24.68\pm0.04$ & $24.10\pm0.04$ & $24.92\pm0.05$ & $24.42\pm0.03$ \\
\end{tabular}
\end{center}
\end{table}
\clearpage
\begin{table}
\begin{center}
Table \thetable---Continued \\
\begin{tabular}{cccccc}
\\
\tableline
\tableline
         &         C3-V14 &         C3-V15 &         C3-V16 &         C3-V17 &         C3-V18 \\
\tableline
HJD$^a$  & F555W          & F555W          & F555W          & F555W          & F555W          \\
273.6254 & $24.75\pm0.04$ & $24.93\pm0.05$ & $24.02\pm0.03$ & $24.57\pm0.04$ & $23.81\pm0.02$ \\
273.6900 & $24.90\pm0.04$ & $24.94\pm0.04$ & $24.40\pm0.03$ & $24.55\pm0.03$ & $23.86\pm0.03$ \\
274.5657 & $24.58\pm0.04$ & $24.71\pm0.04$ & $24.59\pm0.05$ & $24.44\pm0.04$ & $22.98\pm0.01$ \\
274.6302 & $24.28\pm0.03$ & $24.56\pm0.07$ & $24.79\pm0.05$ & $24.29\pm0.03$ & $23.03\pm0.04$ \\
274.9691 & $24.61\pm0.04$ & $24.90\pm0.04$ & $24.56\pm0.04$ & $24.31\pm0.03$ & $23.31\pm0.02$ \\
\end{tabular}
\end{center}
\end{table}
\clearpage
\begin{table}
\begin{center}
Table \thetable---Continued \\
\begin{tabular}{cccccc}
\\
\tableline
\tableline
         &         C3-V19 &         C3-V20 &         C3-V21 &         C3-V22 &         C3-V23 \\
\tableline
HJD$^a$  & F555W          & F555W          & F555W          & F555W          & F555W          \\
273.6254 & $24.72\pm0.04$ & $24.09\pm0.03$ & $25.07\pm0.05$ & $25.15\pm0.06$ & $23.70\pm0.02$ \\
273.6900 & $24.45\pm0.03$ & $24.26\pm0.03$ & $24.92\pm0.04$ & $24.44\pm0.04$ & $23.82\pm0.02$ \\
274.5657 & $24.52\pm0.05$ & $24.52\pm0.04$ & $25.02\pm0.11$ & $25.11\pm0.05$ & $24.44\pm0.03$ \\
274.6302 & $24.82\pm0.04$ & $24.48\pm0.05$ & $25.24\pm0.06$ & $25.16\pm0.05$ & $24.48\pm0.03$ \\
274.9691 & $24.61\pm0.04$ & $24.41\pm0.03$ & $24.87\pm0.06$ & $24.93\pm0.04$ & $24.46\pm0.03$ \\
\end{tabular}
\end{center}
\end{table}
\clearpage
\begin{table}
\begin{center}
Table \thetable---Continued \\
\begin{tabular}{cccccc}
\\
\tableline
\tableline
         &         C3-V24 &         C4-V01 &         C4-V02 &         C4-V03 &         C4-V04 \\
\tableline
HJD$^a$  & F555W          & F555W          & F555W          & F555W          & F555W          \\
273.6254 & $24.82\pm0.04$ & $24.44\pm0.03$ & $24.08\pm0.07$ & $24.51\pm0.04$ & $23.77\pm0.02$ \\
273.6900 & $24.87\pm0.05$ & $24.56\pm0.03$ & $24.44\pm0.03$ & $24.75\pm0.04$ & $23.26\pm0.02$ \\
274.5657 & $24.68\pm0.04$ &     . . .      & $24.55\pm0.04$ & $24.34\pm0.03$ & $23.64\pm0.05$ \\
274.6302 & $24.76\pm0.04$ & $24.47\pm0.03$ & $24.59\pm0.04$ & $24.50\pm0.04$ & $23.93\pm0.03$ \\
274.9691 & $24.80\pm0.05$ & $23.97\pm0.03$ & $24.11\pm0.03$ & $24.82\pm0.05$ &     . . .      \\
\end{tabular}
\end{center}
\end{table}
\clearpage
\begin{table}
\begin{center}
Table \thetable---Continued \\
\begin{tabular}{cccccc}
\\
\tableline
\tableline
         &         C4-V05 &         C4-V06 &         C4-V07 &         C4-V08 &         C4-V09 \\
\tableline
HJD$^a$  & F555W          & F555W          & F555W          & F555W          & F555W          \\
273.6254 &     . . .      & $23.48\pm0.02$ & $24.25\pm0.03$ & $24.66\pm0.06$ & $24.07\pm0.04$ \\
273.6900 & $24.72\pm0.05$ & $23.48\pm0.02$ & $24.38\pm0.03$ & $24.51\pm0.03$ & $24.14\pm0.05$ \\
274.5657 & $24.68\pm0.04$ & $23.89\pm0.03$ & $23.52\pm0.02$ & $25.04\pm0.06$ & $23.65\pm0.03$ \\
274.6302 & $24.68\pm0.06$ & $23.75\pm0.02$ & $23.67\pm0.02$ & $25.07\pm0.05$ & $23.36\pm0.02$ \\
274.9691 & $24.48\pm0.05$ & $23.60\pm0.02$ & $24.18\pm0.03$ & $24.49\pm0.03$ & $23.79\pm0.04$ \\
\end{tabular}
\end{center}
\end{table}
\clearpage
\begin{table}
\begin{center}
Table \thetable---Continued \\
\begin{tabular}{cccccc}
\\
\tableline
\tableline
         &         C4-V10 &         C4-V11 &         C4-V12 &         C4-V13 &         C4-V14 \\
\tableline
HJD$^a$  & F555W          & F555W          & F555W          & F555W          & F555W          \\
273.6254 & $24.52\pm0.03$ & $23.92\pm0.06$ & $24.18\pm0.03$ & $24.42\pm0.10$ & $23.59\pm0.02$ \\
273.6900 & $24.28\pm0.03$ & $24.35\pm0.08$ & $24.32\pm0.04$ &     . . .      & $23.58\pm0.02$ \\
274.5657 & $24.36\pm0.03$ & $25.15\pm0.15$ & $24.96\pm0.05$ & $24.80\pm0.04$ & $23.72\pm0.06$ \\
274.6302 & $24.39\pm0.03$ & $25.16\pm0.11$ & $24.73\pm0.04$ & $24.83\pm0.04$ & $23.56\pm0.02$ \\
274.9691 & $24.73\pm0.07$ & $24.55\pm0.04$ & $24.68\pm0.04$ & $24.27\pm0.03$ & $24.02\pm0.03$ \\
\end{tabular}
\end{center}
\end{table}
\clearpage
\begin{table}
\begin{center}
Table \thetable---Continued \\
\begin{tabular}{cccccc}
\\
\tableline
\tableline
         &         C4-V15 &         C4-V16 &         C4-V17 &         C4-V18 &         C4-V19 \\
\tableline
HJD$^a$  & F555W          & F555W          & F555W          & F555W          & F555W          \\
273.6254 & $24.01\pm0.06$ & $23.79\pm0.02$ & $23.66\pm0.03$ & $23.34\pm0.02$ & $24.36\pm0.03$ \\
273.6900 & $24.12\pm0.06$ & $23.85\pm0.02$ & $23.66\pm0.07$ & $23.45\pm0.03$ & $24.38\pm0.03$ \\
274.5657 & $24.44\pm0.10$ & $23.47\pm0.06$ & $24.77\pm0.07$ & $24.43\pm0.04$ &     . . .      \\
274.6302 &     . . .      & $23.55\pm0.02$ & $24.75\pm0.05$ & $24.41\pm0.04$ & $24.02\pm0.03$ \\
274.9691 & $23.99\pm0.06$ & $23.87\pm0.02$ & $24.01\pm0.05$ & $24.40\pm0.03$ & $24.29\pm0.03$ \\
\end{tabular}
\end{center}
\end{table}
\clearpage
\begin{table}
\begin{center}
Table \thetable---Continued \\
\begin{tabular}{cccccc}
\\
\tableline
\tableline
         &         C4-V20 &         C4-V21 &         C4-V22 &         C4-V23 &         C4-V24 \\
\tableline
HJD$^a$  & F555W          & F555W          & F555W          & F555W          & F555W          \\
273.6254 & $24.49\pm0.03$ & $24.41\pm0.03$ & $24.38\pm0.03$ & $24.49\pm0.07$ & $25.32\pm0.07$ \\
273.6900 & $24.71\pm0.04$ & $24.57\pm0.03$ & $24.55\pm0.03$ & $24.75\pm0.04$ & $25.29\pm0.06$ \\
274.5657 & $24.71\pm0.04$ & $24.02\pm0.03$ & $24.49\pm0.09$ & $24.51\pm0.04$ & $25.03\pm0.05$ \\
274.6302 & $24.77\pm0.04$ & $24.27\pm0.03$ & $24.75\pm0.04$ & $24.54\pm0.05$ & $24.46\pm0.04$ \\
274.9691 & $24.08\pm0.03$ & $24.82\pm0.04$ & $24.13\pm0.03$ & $24.87\pm0.07$ & $25.05\pm0.05$ \\
\end{tabular}
\end{center}
\end{table}
\clearpage
\begin{table}
\begin{center}
Table \thetable---Continued \\
\begin{tabular}{cccccc}
\\
\tableline
\tableline
         &         C4-V25 &         C4-V26 &                  &                  &                  \\
\tableline
HJD$^a$  & F555W          & F555W          &                  &                  &                  \\
273.6254 & $24.70\pm0.04$ & $23.64\pm0.03$ &                  &                  &                  \\
273.6900 &     . . .      & $23.79\pm0.03$ &                  &                  &                  \\
274.5657 & $24.48\pm0.04$ & $23.46\pm0.02$ &                  &                  &                  \\
274.6302 & $24.49\pm0.05$ & $23.50\pm0.02$ &                  &                  &                  \\
274.9691 & $24.76\pm0.04$ & $23.78\pm0.03$ &                  &                  &                  \\
\end{tabular}
\end{center}
\end{table}

\clearpage
\begin{deluxetable}{lcrrrrrccc}
\tablecaption{Sextans A variable stars. \label{tabvars}}
\tablehead{
\colhead{\scriptsize ID} &
\colhead{\scriptsize C} &
\colhead{\scriptsize X} &
\colhead{\scriptsize Y} &
\colhead{\scriptsize $\langle V \rangle$\tablenotemark{a}} &
\colhead{\scriptsize $\langle I \rangle$\tablenotemark{a}} &
\colhead{\scriptsize $\langle V-I \rangle$\tablenotemark{a}} &
\colhead{\scriptsize P} &
\colhead{\scriptsize M\tablenotemark{b}} &
\colhead{\scriptsize Q\tablenotemark{c}}}
\startdata
\scriptsize C1-V01 & \scriptsize 1 & \scriptsize  73.62 & \scriptsize 639.62 & \scriptsize $23.95 \pm 0.10$ & \scriptsize $23.50 \pm 0.05$ & \scriptsize $0.45 \pm 0.11$ & \scriptsize $0.99 \pm 0.04$ & \scriptsize FO & \scriptsize 3 \\
\scriptsize C1-V02 & \scriptsize 1 & \scriptsize  96.66 & \scriptsize 130.15 & \scriptsize $24.52 \pm 0.08$ & \scriptsize $24.04 \pm 0.05$ & \scriptsize $0.48 \pm 0.09$ & \scriptsize $1.24 \pm 0.12$ & \scriptsize FM & \scriptsize 3 \\
\scriptsize C1-V03 & \scriptsize 1 & \scriptsize 203.36 & \scriptsize 660.19 & \scriptsize $24.10 \pm 0.11$ & \scriptsize $23.37 \pm 0.06$ & \scriptsize $0.73 \pm 0.13$ & \scriptsize $1.64 \pm 0.07$ & \scriptsize FM & \scriptsize 4 \\
\scriptsize C1-V04 & \scriptsize 1 & \scriptsize 209.99 & \scriptsize 464.54 & \scriptsize $24.45 \pm 0.09$ & \scriptsize $24.02 \pm 0.06$ & \scriptsize $0.43 \pm 0.11$ & \scriptsize $0.76 \pm 0.04$ & \scriptsize FO & \scriptsize 3 \\
\scriptsize C1-V05 & \scriptsize 1 & \scriptsize 309.49 & \scriptsize 569.75 & \scriptsize $24.48 \pm 0.11$ & \scriptsize $23.90 \pm 0.06$ & \scriptsize $0.58 \pm 0.13$ & \scriptsize $1.24 \pm 0.08$ & \scriptsize FM & \scriptsize 4 \\
\scriptsize C1-V06 & \scriptsize 1 & \scriptsize 372.77 & \scriptsize 143.40 & \scriptsize $24.83 \pm 0.07$ & \scriptsize $24.30 \pm 0.05$ & \scriptsize $0.53 \pm 0.09$ & \scriptsize $0.58 \pm 0.03$ & \scriptsize FO & \scriptsize 4 \\
\scriptsize C1-V07 & \scriptsize 1 & \scriptsize 496.71 & \scriptsize 156.81 & \scriptsize $24.65 \pm 0.11$ & \scriptsize $24.22 \pm 0.07$ & \scriptsize $0.43 \pm 0.13$ & \scriptsize $0.97 \pm 0.04$ & \scriptsize FM & \scriptsize 3 \\
\scriptsize C1-V08 & \scriptsize 1 & \scriptsize 582.38 & \scriptsize 578.70 & \scriptsize $24.73 \pm 0.21$ & \scriptsize $24.18 \pm 0.06$ & \scriptsize $0.55 \pm 0.22$ & \scriptsize $1.08 \pm 0.07$ & \scriptsize FM & \scriptsize 3 \\
\enddata
\tablenotetext{a}{Mean magnitudes are phase-averaged magnitudes.  Uncertainties are formal errors only and thus do not account for uncertainties caused by limited phase coverage.}
\tablenotetext{b}{Modes are ``FM'' (fundamental) and ``FO'' (first overtone).}
\tablenotetext{c}{Variable star quality is based on the cleanness of the light curves, uniqueness of the period, and coherence between F555W and F814W light curves, where 4 is the maximum.}
\end{deluxetable}

\clearpage
\begin{deluxetable}{lcccccr}
\tablecaption{Distance Moduli \label{tabcompare}}
\tablehead{
\colhead{\scriptsize Method} &
\colhead{\scriptsize Sex A} &
\colhead{\scriptsize Leo A} &
\colhead{\scriptsize IC 1613} &
\colhead{\scriptsize SMC} &
\colhead{\scriptsize LMC} &
\colhead{\scriptsize Ref\tablenotemark{a}}}
\startdata
\scriptsize RR Lyrae\tablenotemark{b}        & \scriptsize      . . .       & \scriptsize $24.52 \pm 0.09$ & \scriptsize $24.30 \pm 0.05$ & \scriptsize $18.86 \pm 0.07$ & \scriptsize $18.47 \pm 0.07$ & \scriptsize 1,2,3,4,13 \\
\scriptsize LP Cepheid\tablenotemark{c}      & \scriptsize $25.78 \pm 0.15$ & \scriptsize     . . .        & \scriptsize $24.42 \pm 0.13$ & \scriptsize $18.91 \pm 0.04$ & \scriptsize $18.50 \pm 0.00$ & \scriptsize 5,6,11,14 \\
\scriptsize SP Cepheid (FM)\tablenotemark{d} & \scriptsize $25.64 \pm 0.05$ & \scriptsize $24.61 \pm 0.04$ & \scriptsize $24.42 \pm 0.12$ & \scriptsize $18.88 \pm 0.01$ & \scriptsize $18.30 \pm 0.04$ & \scriptsize 1,2,7,8,9 \\
\scriptsize SP Cepheid (FO)\tablenotemark{d} & \scriptsize $25.58 \pm 0.03$ & \scriptsize $24.49 \pm 0.03$ & \scriptsize $24.38 \pm 0.08$ & \scriptsize $18.88 \pm 0.01$ & \scriptsize $18.32 \pm 0.01$ & \scriptsize 1,2,7,8,9 \\
\scriptsize RGB Tip\tablenotemark{e}         & \scriptsize $25.69 \pm 0.06$ & \scriptsize $24.53 \pm 0.06$\tablenotemark{f} & \scriptsize $24.28 \pm 0.05$\tablenotemark{g} & \scriptsize $18.88 \pm 0.04$\tablenotemark{h} & \scriptsize $18.52 \pm 0.09$ & \scriptsize 7,10,11 \\
\scriptsize Red Clump\tablenotemark{i}       & \scriptsize $25.51 \pm 0.15$ & \scriptsize $24.43 \pm 0.15$\tablenotemark{f} & \scriptsize $24.27 \pm 0.08$\tablenotemark{g} & \scriptsize $18.85 \pm 0.06$ & \scriptsize $18.46 \pm 0.11$ & \scriptsize 7,12 \\
\enddata
\end{deluxetable}

\clearpage
\begin{deluxetable}{ccccc}
\tablecaption{Distance Moduli Relative to SMC \label{tabrelative}}
\tablehead{
\colhead{Galaxy} &
\colhead{Sex A} &
\colhead{Leo A} &
\colhead{IC 1613} &
\colhead{LMC}}
\startdata
$(m-M)_0-(m-M)_{0,SMC}$ & $6.75 \pm 0.04$ &  $5.65 \pm 0.04$ &  $5.44 \pm 0.04$ & $-0.37 \pm 0.05$ \\
\enddata
\end{deluxetable}

\end{document}